\documentclass[review]{elsarticle}
\usepackage{adjustbox}
\usepackage{transparent}
\usepackage{graphicx}
\usepackage{float}
\usepackage{subfig}
\usepackage{xcolor}
\usepackage{multirow}
\usepackage{array}
\usepackage{algpseudocode}
\usepackage[ruled,vlined]{algorithm2e}

\SetCommentSty{mycommfont}
\usepackage{lineno,hyperref,xcolor}

\usepackage{tikz,amsmath}
\usetikzlibrary{matrix,calc}
\usetikzlibrary{shapes.multipart}
\usetikzlibrary{shapes.geometric,arrows} % to implement the corresponding shapes and arrows, are used.  
\newcommand*{\green}[1]{
    \protect
  \begin{tikzpicture}[scale=#1]
    \protect \draw[draw=green!65, fill=green!70] (0,0) -- (1,0) -- (0.5,0.87) -- cycle;
  \end{tikzpicture}
}

\newcommand*{\invgreen}[1]{
    \protect
  \begin{tikzpicture}[scale=#1]
    \protect \draw[draw=green!65, fill=green!70] (0,0) -- (1,0) -- (0.5,-0.87) -- cycle;
  \end{tikzpicture}
}

\newcommand*{\red}[1]{%
    \protect
  \begin{tikzpicture}[scale=#1]
    \protect \draw[draw=red!55, fill=red!70] (0,0) -- (1,0) -- (0.5,-0.87) -- cycle;
  \end{tikzpicture}%
}

\newcommand*{\invred}[1]{%
    \protect
  \begin{tikzpicture}[scale=#1]
    \protect \draw[draw=red!55, fill=red!70] (0,0) -- (1,0) -- (0.5,0.87) -- cycle;
  \end{tikzpicture}%
}

\newcommand{\up}[1]{\begin{centering}{\green{0.25}{\hspace{1mm}#1}}\end{centering}}
\newcommand{\down}[1]{\begin{centering}{\red{0.25}{\hspace{1mm}#1}}\end{centering}}

\newcommand{\invdownp}[1]{\invred{0.25}{\hspace{1mm}#1}\%}
\newcommand{\upp}[1]{\green{0.25}{\hspace{1mm}#1}\%}
\newcommand{\invupp}[1]{\invgreen{0.25}{\hspace{1mm}#1}\%}
\newcommand{\downp}[1]{\red{0.25}{\hspace{1mm}#1}\%}

\SetKwInput{KwInput}{Input}                % Set the Input
\SetKwInput{KwOutput}{Output}              % set the Output
\usepackage{color, soul}
\sethlcolor{red}
\journal{Computers in Biology and Medicine}

\begin{document}

\begin{frontmatter}

\title{Integrating Edges into U-Net Models with Explainable Activation Maps for Brain Tumor Segmentation using MR Images}

%% Group authors per affiliation:

\author[1,2]{Subin Sahayam}
\corref{cor1}
\ead{subinsahayamm@snuchennai.edu.in, coe18d001@iiitdm.ac.in}

\address[1]{Department of Computer Science and Engineering, Shiv Nadar University Chennai, Kalavakkam - 603110, Tamil Nadu, India}

\address[2]{Department of Computer Science and Engineering, Indian Institute of Information Technology Design and Manufacturing Kancheepuram, Chennai - 600127, Tamil Nadu, India}

\author[2]{Umarani Jayaraman}
\ead{umarani@iiitdm.ac.in}

\cortext[cor1]{Corresponding author}

\begin{abstract}
Manual delineation of tumor regions from magnetic resonance (MR) images is time-consuming, requires an expert, and is prone to human error. In recent years, deep learning models have been the go-to approach for the segmentation of brain tumors.  U-Net and its' variants for semantic segmentation of medical images have achieved good results in the literature. However, U-Net and its' variants tend to over-segment tumor regions and may not accurately segment the tumor edges. The edges of the tumor are as important as the tumor regions for accurate diagnosis, surgical precision, and treatment planning. In the proposed work, the authors aim to extract edges from the ground truth using a derivative-like filter followed by edge reconstruction to obtain an edge ground truth in addition to the brain tumor ground truth. Utilizing both ground truths, the author studies several U-Net and its' variant architectures with and without tumor edges ground truth as a target along with the tumor ground truth for brain tumor segmentation. The author used the BraTS2020 benchmark dataset to perform the study and the results are tabulated for the dice and Hausdorff95 metrics. The mean and median metrics are calculated for the whole tumor (WT), tumor core (TC), and enhancing tumor (ET) regions. Compared to the baseline U-Net and its variants, the models that learned edges along with the tumor regions performed well in the enhancing and core tumor regions in both training and validation datasets. The improved performance of edge-trained models trained on baseline models like U-Net and V-Net achieved performance similar to baseline state-of-the-art models like Swin U-Net and hybrid MR-U-Net.  The edge-target trained models are capable of generating edge maps that can be useful for treatment planning. Additionally, for further explainability of the results, the activation map generated by the hybrid MR-U-Net has been studied.
\end{abstract}

\begin{keyword}
 BraTS2020 \sep Boundary Aware Model \sep Tumor Boundary \sep U-Net \sep Activation map
\end{keyword}

\end{frontmatter}
\section{Introduction}
\label{chap1}
Gliomas are a group of primary brain tumors that arise from glial cells. It is a significant cause of morbidity and mortality worldwide. According to the latest statistics by the Central Brain Tumor Registry of the United States, about 84,264 deaths occurred due to malignant brain and central nervous system (CNS) tumors between 2015 and 2019. Also, 24\% of primary brain tumors diagnosed in the US were gliomas \cite{ostrom2022cbtrus}. 

Accurate diagnosis of gliomas is critical for treatment planning and improved patient outcomes. Magnetic resonance imaging (MRI) is the modality of choice for evaluating gliomas due to its high sensitivity and ability to provide detailed anatomical information. The diagnosis from MR images involves a comprehensive analysis of various parameters, including tumor location, size, morphology, and enhancement pattern. The combination of MRI modalities like fluid attenuation inversion recovery (FLAIR), T1-weighted (T1), contrast-enhanced T1-weighted (T1ce), diffusion-weighted imaging (DWI), perfusion-weighted imaging (PWI) and T2-weighted (T2) can provide a detailed assessment of glioma features, including tumor grade, invasiveness, and response to therapy \cite{mo2020multimodal, sun2019drrnet}. However, manual segmentation is a time-consuming and labor-intensive task that requires extensive expertise and is subject to inter-observer and intra-observer variability. The dice score for manual segmentation generally ranges between  74–85 \% \cite{menze2014multimodal}. Automatic segmentation of gliomas from MR images using deep learning-based methods has shown great promise in recent years. Reliable and efficient segmentation methods for gliomas are essential for improving patient outcomes and accelerating research into the disease \cite{agravat20213d, greene2018behavioral}.

\begin{figure}[htb]
\captionsetup[subfigure]{justification=centering}
	\centering
	\subfloat[FLAIR Image]{{\includegraphics[width=0.21\linewidth]{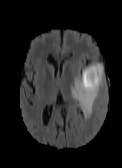} }}
	\quad
	\subfloat[T1CE Image]{{\includegraphics[width=0.21\linewidth]{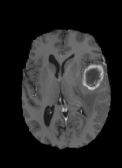} }}
	\quad
	\subfloat[T2 Image]{{\includegraphics[width=0.21\linewidth]{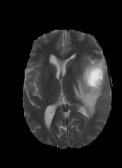} }}
	\quad
	\subfloat[Ground Truth]{{\includegraphics[width=0.21\linewidth]{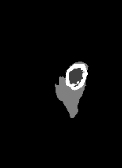} }}
	
	\caption{Shows a sample 2D axial input MR Images (a-c) and the corresponding ground truth (d). In the ground truth, the white region corresponds to the enhancing tumor (ET), the dark grey region corresponds to the necrotic core region and non-enhancing tumor (NCR/NET), and the light grey area represents the edema region (ED) \cite{sahayam2022brain}.}
 \label{mr_input_and_output1}
\end{figure}

The Brain Tumor Segmentation (BraTS) \cite{menze2014multimodal, Bakas2017} challenge is an annual international competition that evaluates state-of-the-art algorithms for brain tumor segmentation from MR images. The challenge provides a standardized dataset of multi-modal MR scans, including T1, FLAIR, T1CE, and T2 MR images. These images help identify and mark various regions affected by tumors in the brain. The peritumoral edema (ED) region appears with relatively high intensity in the FLAIR image. Enhancing tumor (ET) regions appear hyperintense in the T1CE image compared to the healthy and non-enhancing tumor volumes. The T2 image shows the necrotic core and non-enhancing tumor (NCR/NET) region \cite{bakas2018identifying}. A sample 2D axial slice of FLAIR, T1CE, T2 MR, and the corresponding ground-truth image is shown in Figure \ref{mr_input_and_output1}.

 Deep learning models have recently become the go-to semantic segmentation technique. Some of the popular semantic segmentation models for brain tumor segmentation are U-Net \cite{ronneberger2015u}, V-Net \cite{milletari2016v}, Attention U-Net \cite{oktay2018attention}, U-Net 3+ \cite{huang2020unet}, Hybrid MR-U-Net \cite{sahayam2022brain} and Swin-U-Net \cite{cao2023swin}.  Tumor edge information is equally important as the tumor region for segmentation. The majority of proposed works in the literature focus on model-based improvements and a few focus on the fusion of multi-modal MRI inputs \cite{mo2020multimodal, zhang2020exploring} for semantic segmentation. Very few works extract edges and try to bring edge information into their models to improve the overall segmentation result \cite{jiang2021novel, zhu2023brain}. In this work, the author aims to take a data-oriented approach by fixing popular semantic segmentation models and trying to directly learn tumor edges as direct targets along with the other tumor regions.

 The contribution of the work is as follows,

 \begin{enumerate}
     \item Identify edges from the ground truth using a 3D Laplacian-like filter and utilizing the ground truth as a reference image to extract tumor edges.
     \item Learning the edges and the tumor regions as targets using various deep learning models like U-Net \cite{ronneberger2015u}, V-Net \cite{milletari2016v}, Attention U-Net \cite{oktay2018attention}, U-Net 3+ \cite{huang2020unet}, and Swin-U-Net \cite{cao2023swin}.
     \item A custom focal loss has been proposed to give higher weights to edges than tumor  regions.
     \item Studying the performance of the segmented tumor on the BraTS2020\cite{menze2014multimodal} training and validation dataset.
     \item Analyzing activation maps of the last layer for explainability of models with and without edges as targets along with tumor regions.
 \end{enumerate}

The remaining part of the paper is organized as follows. The next section \ref{chap2} discusses several works relating to brain tumor segmentation in the literature. Section \ref{chap3} details the proposed methodology and the individual blocks in the workflow. The following section \ref{chap4} briefs about the dataset, implementation details, and the experiments carried out to evaluate the proposed workflow.  Finally, section \ref{chap6} summarizes the work, discusses the important points, and suggests future directions.

\section{Related Work}
\label{chap2}
The brain tumor segmentation challenge \cite{bakas2018identifying} (BraTS) from 2012 sparked waves of interest in segmenting tumors from brain MRI. The yearly challenge helped gather multi-institutional data, data standardization, and bench-marking \cite{menze2014multimodal}. The early approaches focused on
 image processing techniques like thresholding, probabilistic methods, and clustering. In \cite{njeh20153d}, Njeh et al. proposed a graph cut distribution matching method to automatically segment edema regions. Abdel et al. \cite{abdel2015brain} proposed a hybrid clustering method using k-means integrated with fuzzy c-means to find the initial brain tumor area. They followed it up with thresholding and level-set segmentation stages to obtain the segmented brain tumors. Both authors validated their methodology on the BraTS2012 dataset. Dvorak et al. \cite{dvorak2015automated} utilized multi-resolution symmetric analysis and Otsu's thresholding on axial and coronal planes. The left symmetry information in both planes has been used as prior knowledge to detect tumors from T2-weighted images and the threshold has been used to segment the same. These methods generally tend to be computationally fast but they cannot generalize to a large variety of MR images for brain tumor segmentation.
 
In recent years, machine learning and deep learning techniques have performed well and given state-of-the-art results. Pinto et al. \cite{pinto2018hierarchical} have proposed a hierarchical brain tumor segmentation workflow that utilizes extremely randomized trees for segmentation. These trees are provided with appearance and context-based features for accurate segmentation. The features are obtained through several non-linear transformations. Havaei et al. \cite{havaei2017brain} proposed several cascade-style deep convolutional neural networks for the segmentation of brain tumors. The authors claim that their methodology learns both local as well as global contextual features. The authors showed their results on the BraTS2013 dataset. The author \cite{wang20213d} proposes an enhanced version of dense convolutional neural networks for the task of brain tumor segmentation. Additionally, they have utilized the super-resolution concept for image reconstruction to obtain quality segmentation results. Kamnitsas et al. \cite{kamnitsas2017efficient} proposed a 3D convolutional neural network for brain tumor segmentation. They also obtain both local and global contextual information, they proposed a dual pathway architecture with multi-scale filters. They removed false positives by applying a conditional random field over the obtained results. Their methodology obtained top performance on the BraTS2015 and ISLES2015 datasets.

The drastic improvements in semantic segmentation started with the introduction of the U-Net model \cite{ronneberger2015u}. Feng et al. \cite{feng2020brain} proposed an ensemble of 3D U-Nets with various hyperparameter settings. They build an ensemble model with different hyperparameters to reduce random errors at different hyperparameter settings. Kamnitsas et al \cite{kamnitsas2018ensembles} proposed an ensemble model named Ensembles of Multiple Models and Architectures (EMMA) for brain tumor segmentation. The model used other top-performing segmentation models like DeepMedic \cite{kamnitsas2016deepmedic}, FCN \cite{long2015fully} and U-Net \cite{ronneberger2015u} to produce a target segmented image. Myroenko \cite{myronenko20193d} proposed a U-Net with an additional variational autoencoder branch. The variational autoencoder part aims at the reconstruction of the input image from the bottleneck region of the U-Net thereby building robust features useful for obtaining proper segmentation at the U-Net segmentation layer. Isensee et al. \cite{isensee2019no} demonstrated that a U-Net is difficult to beat with proper hyperparameter tuning, additional input data, dice loss, and other minor changes. 

Most works in the literature so far have focused on changes to the model architecture or hyperparameter tuning. Also, when it comes to brain tumors, the edge prediction of the tumor is as important as the tumor regions themselves. Edges of the tumor regions are difficult to delineate as the transition from healthy tissue to brain tissue regions is usually smooth in nature \cite{bakas2018identifying}. In recent years, there have been efforts to bridge the gap by designing models that are aware of the edges \cite{liu2022shape}, and models in which the edge information is infused during the learning process \cite{zhu2023brain, jiang2021novel}. The proposed models however are not applicable to other models without major changes to the base model. The simplest way to learn edges would be to directly learn the edges along with the tumor segments as targets when training a model. The approach can be applied to any deep learning model with minimal change to the architecture. To the authors' knowledge, such an endeavor has not been carried out in the literature and the authors feel that such an exercise is worth exploring.

\section{Proposed Methodology}
\label{chap3}
\begin{figure}[htb]
	\centering
	\includegraphics[width=\linewidth]{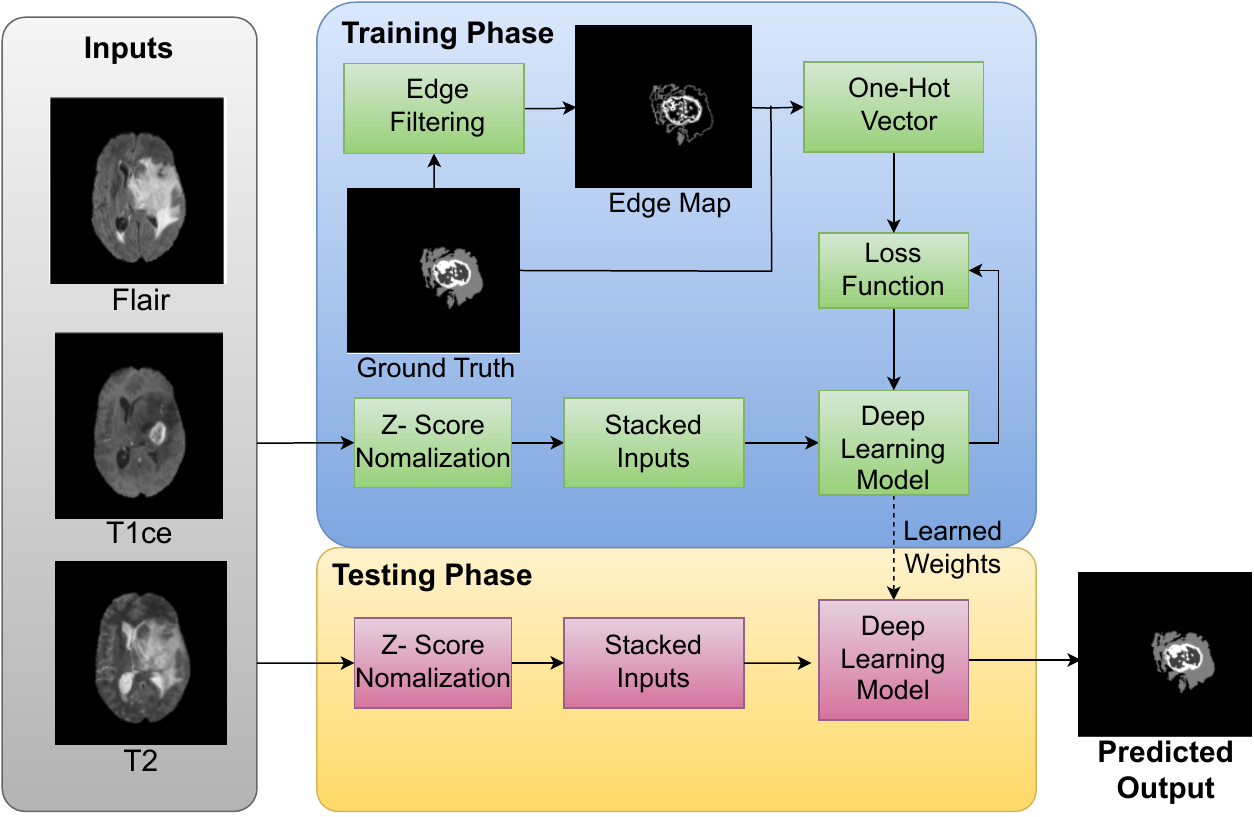}
	\caption{Shows the flow diagram of the proposed methodology for training U-Net models with the ground truth and the edges as targets.}
	\label{block}
\end{figure}

The workflow of the proposed methodology is shown in Figure \ref{block}. The workflow can be broken down into Z-score normalization, edge extraction, one-hot representation, and training deep learning models. The work utilizes FLAIR, T2, and T1CE 3D MRI modalities and the ground truth for segmentation.

\subsection{Z-Score Normalization}
\label{norm}
MR images suffer from intensity inhomogeneity, where image pixel intensities vary due to several factors, thus reducing the contrast between different objects in the image. The variation is observed across 2D slices in any given 3D MR image \cite{ganzetti2016quantitative}. Especially, when comparing MR images of different patients or MR images of different modalities for the same patient, the mean intensity and its standard deviation can vary drastically \cite{ioffe2015batch}. It makes it difficult to learn patterns across MR images acquired with MR scanners of the same or different configurations and scanning procedures.  Thus, it affects the model's ability to generalize.

\begin{figure}[htb]
\captionsetup[subfigure]{justification=centering}
	\centering
	\subfloat[2D Axial slice of FLAIR MRI before normalization]{{\includegraphics[width=0.3\linewidth,height=0.25\linewidth]{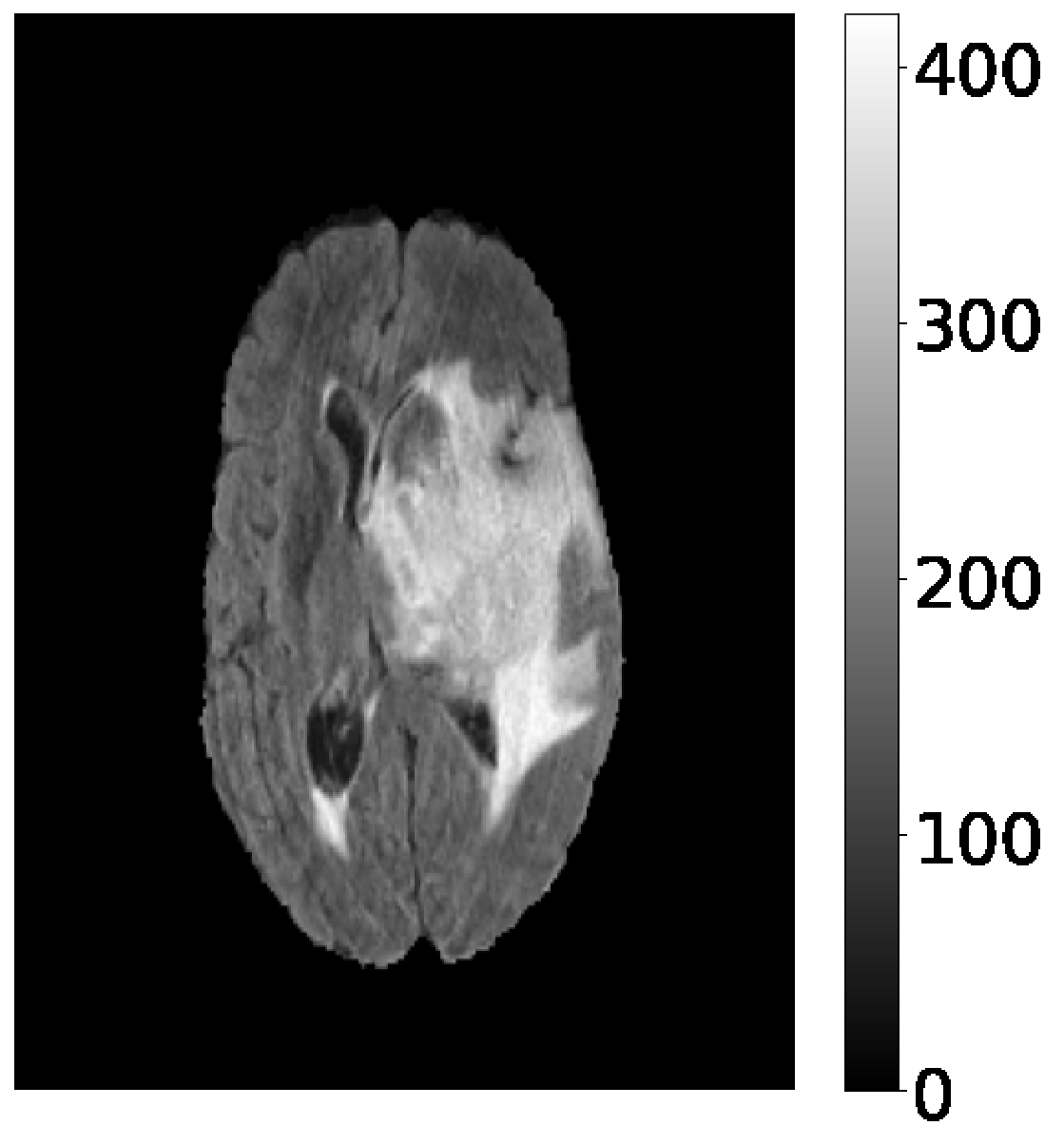} }}
	\quad
	\subfloat[2D Axial slice of T1CE MRI before normalization]{{\includegraphics[width=0.3\linewidth,height=0.25\linewidth]{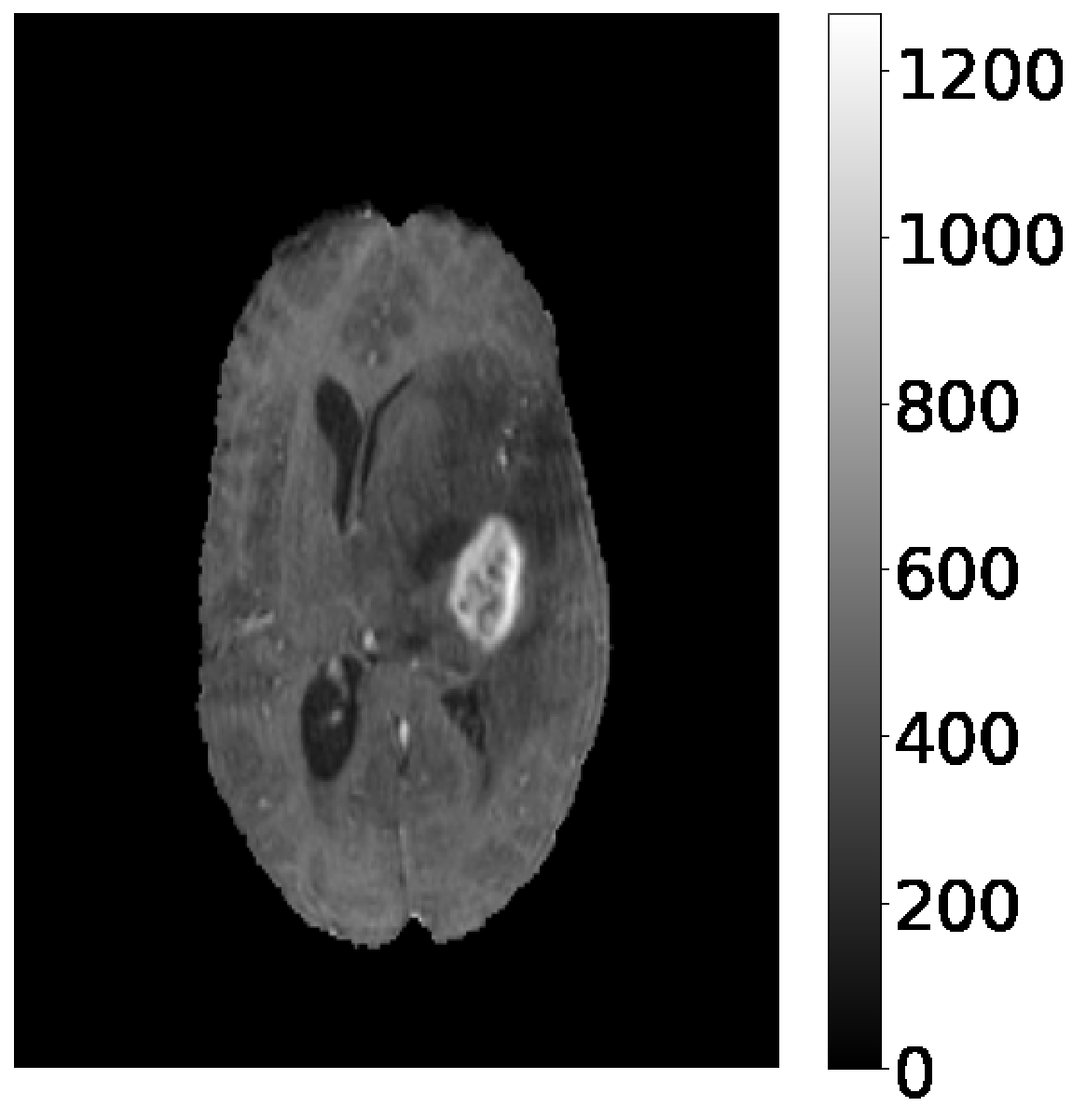} }}
	\quad
	\subfloat[2D Axial slice of T2 MRI before normalization]{{\includegraphics[width=0.3\linewidth,height=0.25\linewidth]{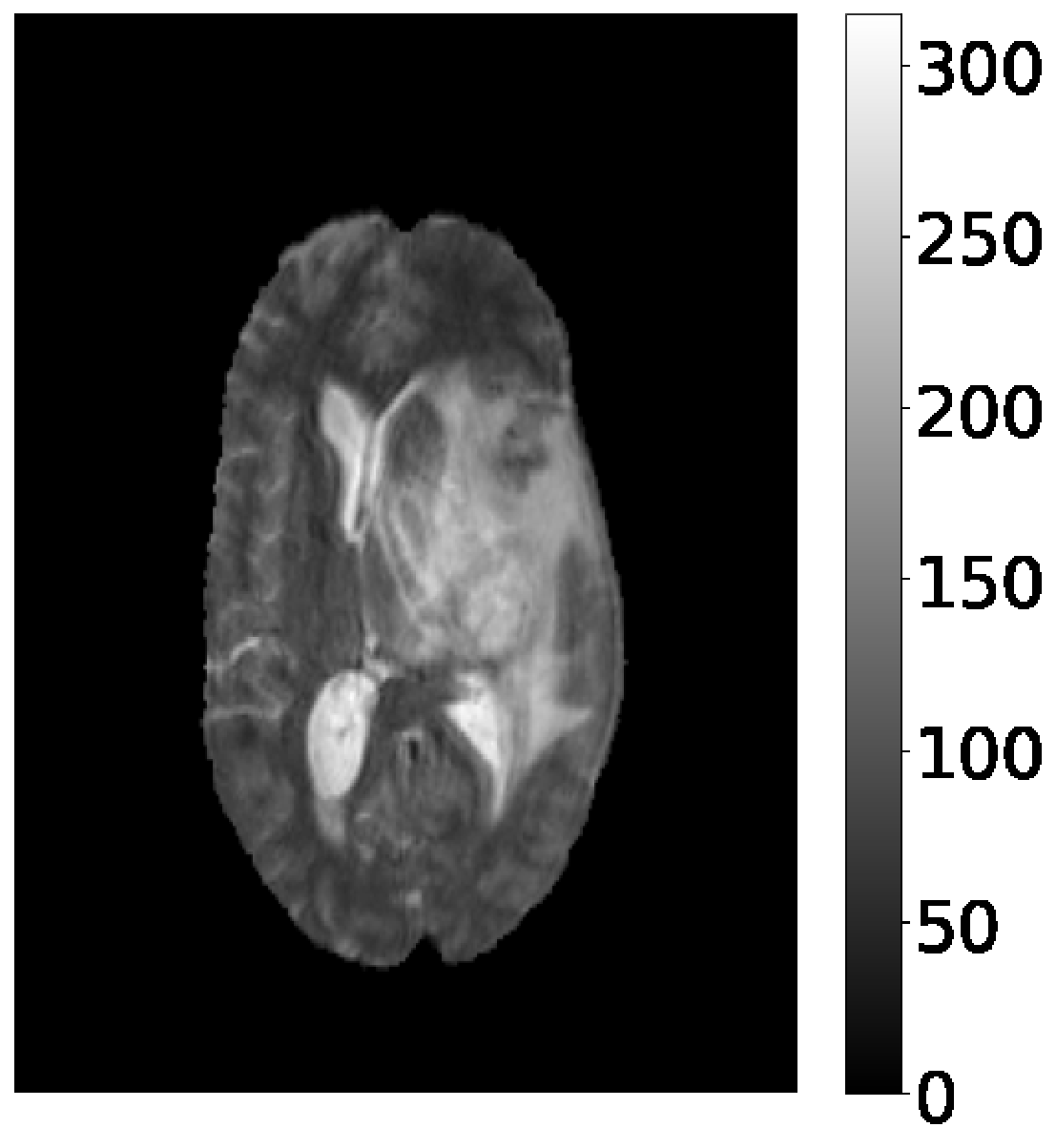} }}
	\quad
	\subfloat[2D Axial slice of FLAIR MRI after normalization]{{\includegraphics[width=0.3\linewidth,height=0.25\linewidth]{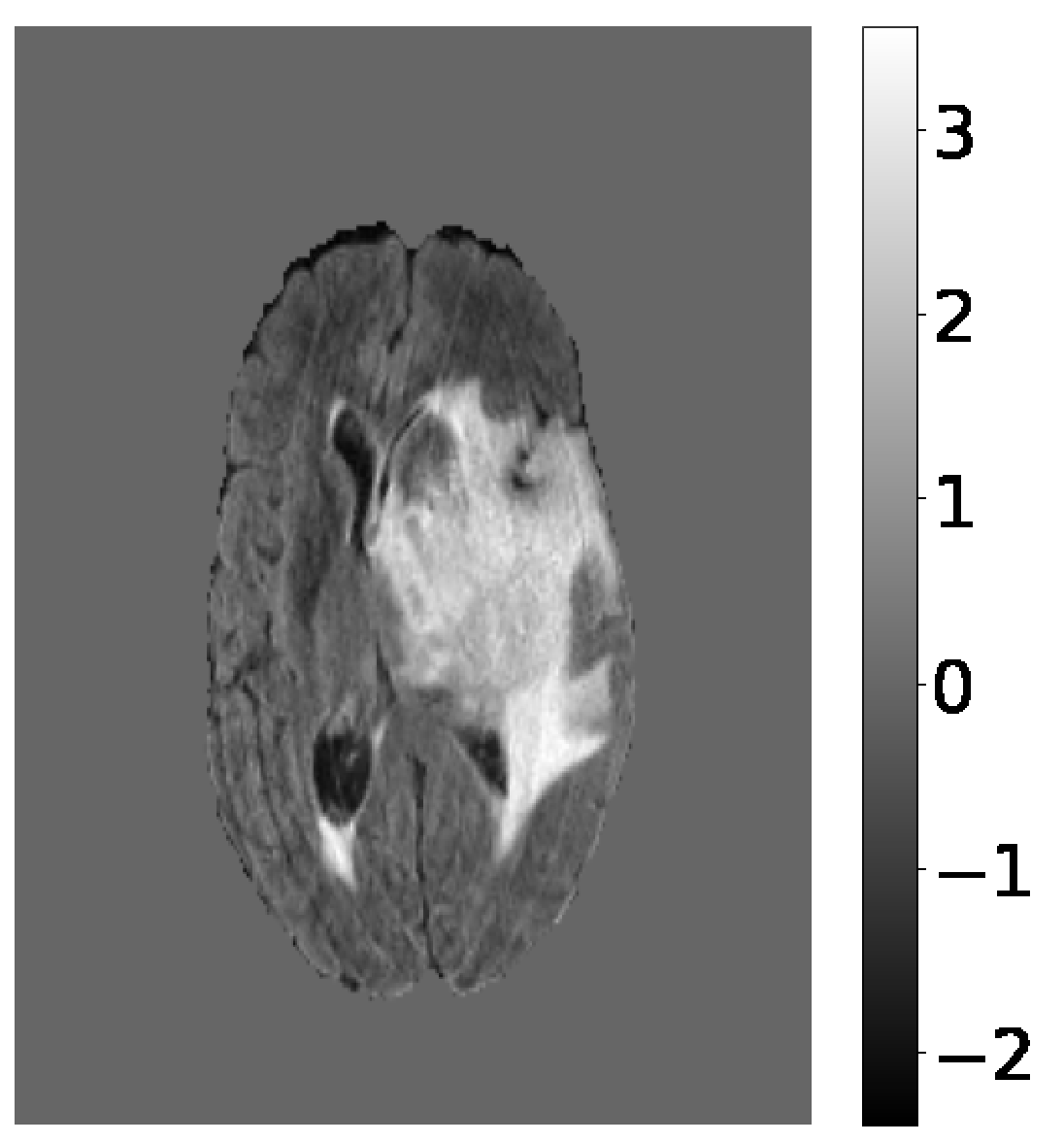} }}
	\quad
	\subfloat[2D Axial slice of T1CE MRI after normalization]{{\includegraphics[width=0.3\linewidth,height=0.25\linewidth]{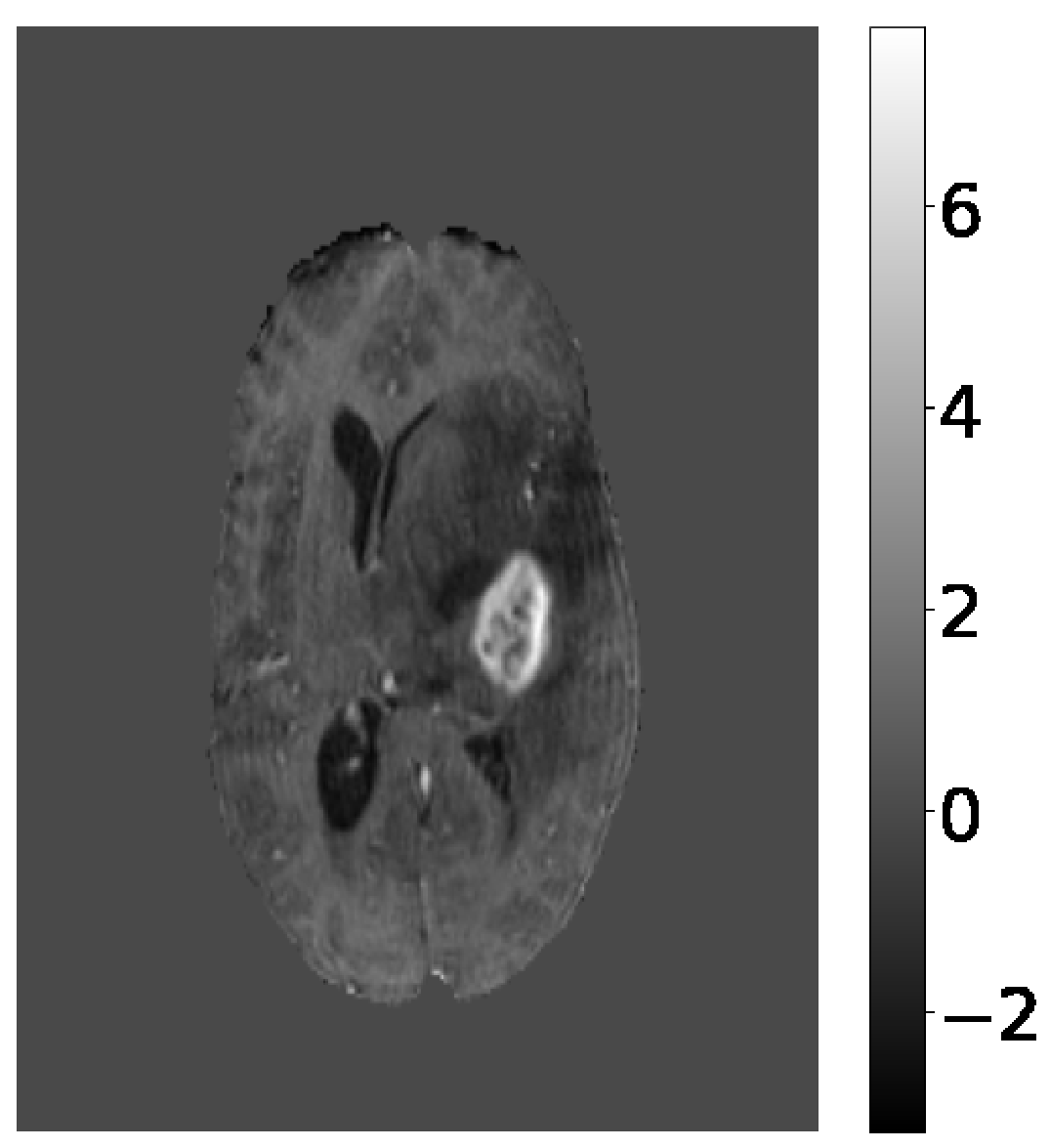} }}
	\quad
	\subfloat[2D Axial slice of T2 MRI after normalization]{{\includegraphics[width=0.3\linewidth,height=0.25\linewidth]{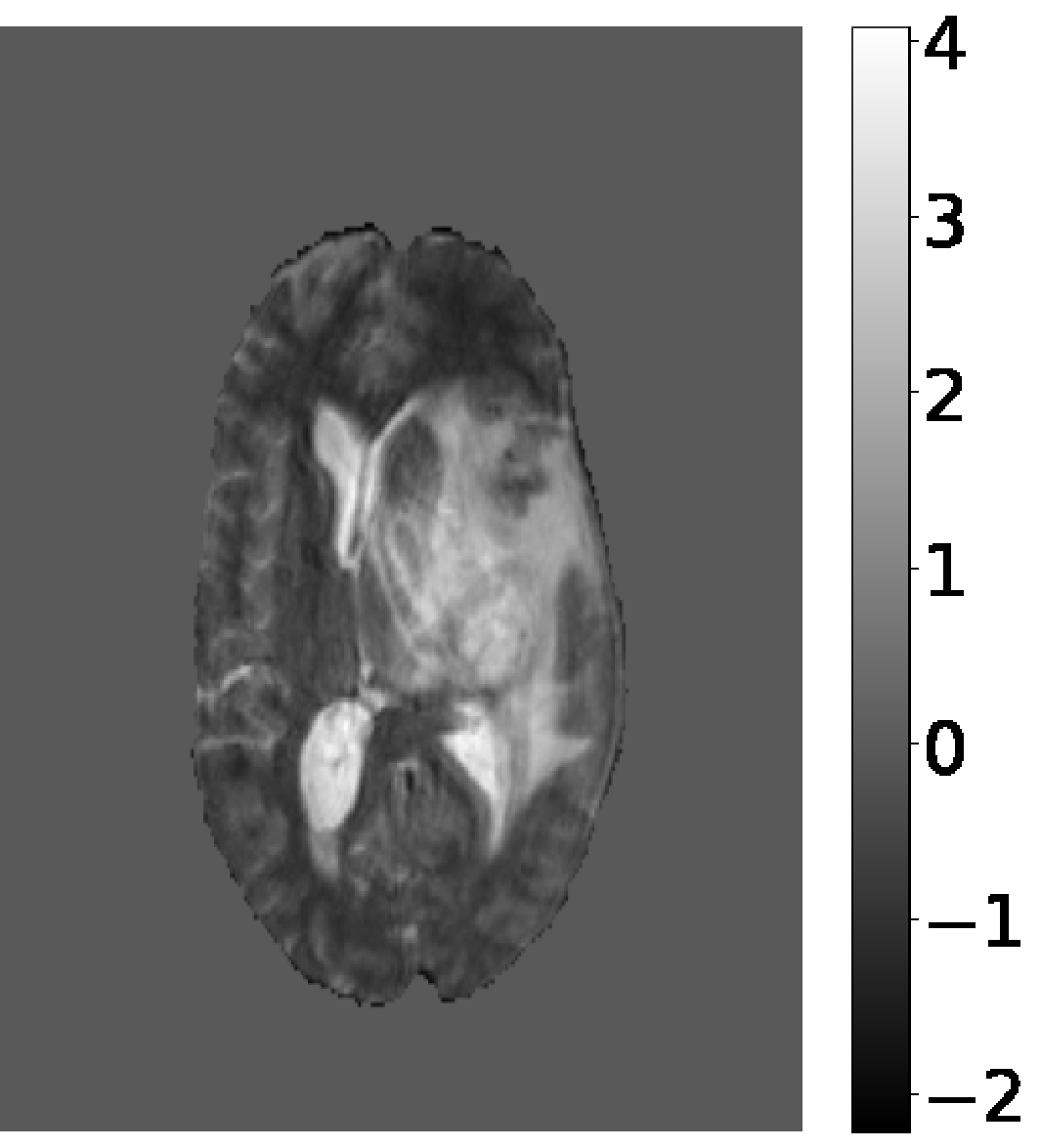} }}
	\caption{Shows the sample set of 2D axial slice input images (FLAIR, T1ce, T2) before (a-c) and the corresponding MR images (FLAIR, T1ce, T2) after (d-f) Z-score normalization for a patient in BraTS2020 dataset \cite{sahayam2022brain}}

\label{fig:mr_input_and_output1}
\end{figure}
 The same Z-score normalization algorithm used in \cite{sahayam2022brain} has been used in the proposed work. Before normalization, each modality can have a varying intensity distribution. For example, Figure \ref{fig:mr_input_and_output1} shows a 2D axial slice of FLAIR, T1CE, and T2 MRI before and after normalization. Before normalization, the FLAIR image intensity range is approximately between 0 to 400, T1CE is between 0 to 1200, and T2 is between 0 to 300 as shown in Figure \ref{fig:mr_input_and_output1} (a-c). The normalization approach is aimed at standardizing the intensity distribution of the brain regions across all three modalities, such that the mean intensity value in the brain region is 0 and the standard deviation is 1. After normalization, the intensity range for FLAIR is approximately between -2 to 3, T1CE is between -2 to 6, and -2 to 4 as shown in Figure \ref{fig:mr_input_and_output1} (d-e). The mean intensity value at the brain region of each of the modalities after normalization would be 0 and the standard deviation would be 1.
 
\subsection{Edge Extraction}
The boundaries between tumor regions are smooth and difficult to differentiate in an MR image. It leads to segmentation predictions with over-segmentation near the boundaries. There are several ways to infuse boundary information into a trained model \cite{jiang2021novel, zhu2023brain}. One of the simplest ways is to directly learn the target boundaries along with the tumor ground truth. Datasets in general only provide target tumor ground truth and not tumor edges. So, the edges need to be extracted from the ground truth images.
\begin{figure}[hb]
     \centering
\[\begin{bmatrix}

\begin{bmatrix}
-1 & -1 & -1 \\
-1 & -1 & -1 \\
-1 & -1 & -1
\end{bmatrix}
&
\begin{bmatrix}
-1 & -1 & -1 \\
-1 & 26 & -1 \\
-1 & -1 & -1
\end{bmatrix}
&
\begin{bmatrix}
-1 & -1 & -1 \\
-1 & -1 & -1 \\
-1 & -1 & -1
\end{bmatrix}
\end{bmatrix}\]

\caption{3D filter used to convolve over the ground truth to obtain the 3D edge mask.}
     \label{3D_laplacian}
\end{figure}

\begin{figure}[htb]
\centering
	\captionsetup[subfigure]{justification=centering}
	\subfloat[2D Axial slice of Ground Truth MRI before boundary extraction]{{\includegraphics[width=0.3\linewidth,height=0.28\linewidth]{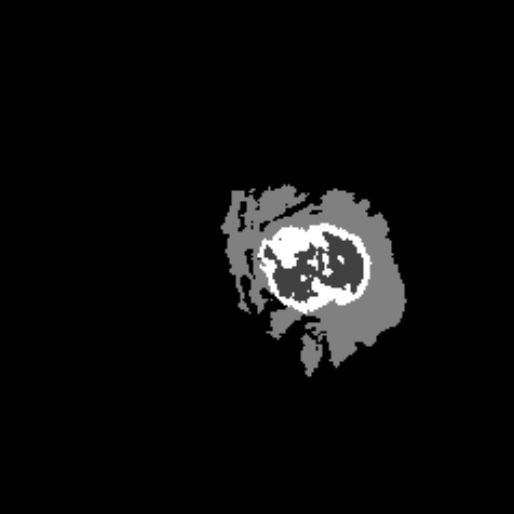} }}
	\quad
	\subfloat[2D Axial slice of Ground Truth MRI after boundary extraction]{{\includegraphics[width=0.3\linewidth,height=0.28\linewidth]{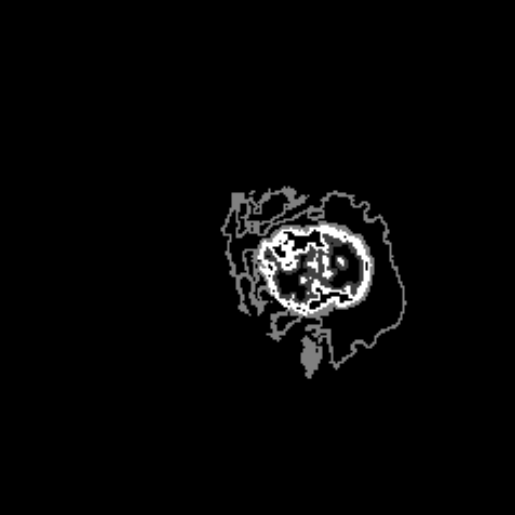} }}
	\caption{Edge mask generated using the convolution operation on the ground truth segmentation mask of a patient followed by edge reconstruction.}
	\label{edge_mask}
\end{figure}

The Laplacian operator \cite{gonzalez2009digital} is a second derivative operator that is used to detect the zero-crossings of image intensity and often yields better edge-detection results. The filter is ineffective in noisy images, but the ground-truth MR image is free from noise and so can be used for extraction of edges. As MR images are 3D images, a 3D Laplacian-like filter is used to extract the edges from MR modalities as shown in \ref{3D_laplacian}. On applying the filter, the region immediately following an edge will have a negative value \cite{naser2019three}. To handle the negative values, the ground truth is used as a reference image, and the edge image is reconstructed as mentioned in the Algorithm \ref{alg:edge}.
\begin{algorithm}[hbt]
\DontPrintSemicolon
  \KwInput{Zero Padded 3D Image $I_{in}$, 3D Laplacian-like filter $\mathcal{F}$.}
  \KwOutput{3D Edge Extracted Image $I_{out}$}
  \For{ $\forall (x, y, z)  \in I_{in}$}{
    \For{$\forall (i, j, k)  \in \mathcal{F}$}{
        \tcc{Convolution Operation of Filter $\mathcal{F}$ Over Input $I_{in}$}
        \tcc{$a, b, c$ are the Filter Size of Filter $\mathcal{F}$ by $2$}
        $T(x,y,z) \gets \sum_{i=-a}^{a}\sum_{j=-b}^{b}\sum_{k=-c}^{c} F(i,j,k) * I_{in}(x+i,y+j,z+k)$\\
    }
  }
  \tcc{Initialize $I_{out}$ of dimension $(x, y, z)$ to $0$}
  $I_{out} \gets 0$\\
  \For{ $\forall (x, y, z)  \in I_{in}$}{
  \tcc{For all non-zero values in $T(x,y,z)$, replace the 0 in $I_{out}(x, y, z)$ with $I_{in}(x, y, z)$ }
    \If{$T(x,y,z) \ne 0$}{
            $I_{out}(x, y, z) \gets I_{in}(x, y, z)$ \\
        }
    }
    \Return{$I_{out}$}
\caption{Extraction of Edges from Ground Truth}
\label{alg:edge}
\end{algorithm}
A sample 2D axial slice of the ground truth image and the corresponding extracted edges are shown in Figure \ref{edge_mask}.

\subsection{One-Hot Representation}
\begin{figure}[hbt]
\captionsetup[subfigure]{justification=centering}
	\centering
	\subfloat[GroundTruth]{{\includegraphics[width=0.2\linewidth,height=0.2\linewidth]{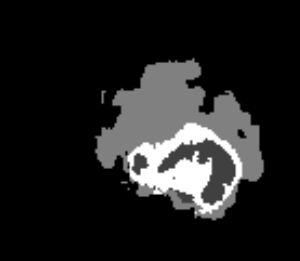} }}
        \quad  
        \subfloat[Edema One-Hot 2D MR Slice]{{\includegraphics[width=0.2\linewidth,height=0.2\linewidth]{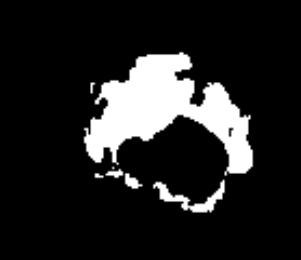} }}
        \quad
	\subfloat[Core One-Hot 2D MR Slice]{{\includegraphics[width=0.2\linewidth,height=0.2\linewidth]{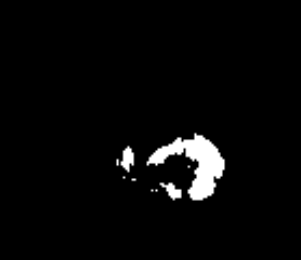} }}
	\quad
     \subfloat[Enhancing One-Hot 2D MR Slice]{{\includegraphics[width=0.2\linewidth,height=0.2\linewidth]{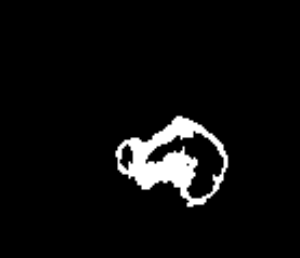} }}
        \quad
        \subfloat[Background One-Hot 2D MR Slice]{\fbox{{\includegraphics[width=0.2\linewidth,height=0.2\linewidth]{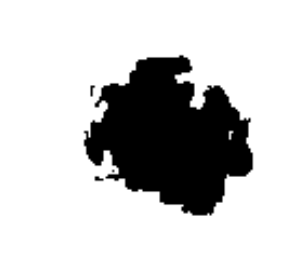} }}}
	\quad
 \subfloat[Edema Edge One-Hot 2D MR Slice]{{\includegraphics[width=0.2\linewidth,height=0.2\linewidth]{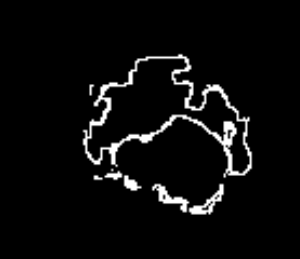} }}
	\quad
     \subfloat[Core Edge One-Hot 2D MR Slice]{{\includegraphics[width=0.2\linewidth,height=0.2\linewidth]{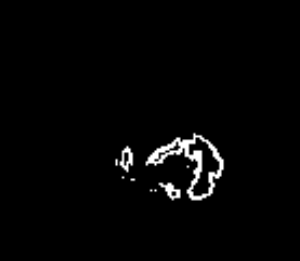} }}
    	\quad
     \subfloat[Enhancing One-Hot 2D MR Slice]{{\includegraphics[width=0.2\linewidth,height=0.2\linewidth]{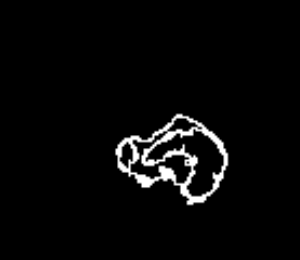} }}
	
	\caption{Shows the sample set of 2D MR images with the ground truth and the corresponding one-hot representation of each region and edge region.}

\label{mul}
\end{figure}

For multiple segmentation, deep learning models require target one-hot encoding. One-hot encoding converts a 1D tensor of size $n$ with $c$ target classes into a 2D tensor of size $n x c$ with $1$ in each row representing one among the $c$ classes and the other values in the row are marked as $0$ \cite{rodriguez2018beyond}. MRI images are 3D images. So, for a 3D target tensor of dimension, $(x,y,z)$ with $c$ classes, a one-hot encoding will result in a matrix with $1$ and $0$ of dimension $(x,y,z,c)$. A 2D slice of the coronal plane is shown for a patient in Figure \ref{mul} (a). The white region corresponds to the enhancing tumor (ET), the dark grey region corresponds to the necrosis and non-enhancing tumor (NCR/NET), and the light grey area represents the edema region (ED). The proposed method utilizes both the ground truth and target edges and so, a one-hot vector should be generated for both of the targets. For the ground truth Figure \ref{mul} (a), the corresponding one-hot vectors for the target tumor region, the background region, and the edges are shown in Figure \ref{mul} (b) - (h).

\subsection{Training Deep Learning Models}
Deep learning models for semantic segmentation have been trained using the proposed workflow. Some of the popular semantic segmentation models used to study the proposed workflow are U-Net \cite{ronneberger2015u}, V-Net \cite{milletari2016v}, Attention U-Net \cite{oktay2018attention}, U-Net 3+ \cite{huang2020unet}, and Swin-U-Net \cite{cao2023swin}. A deep learning model, Hybrid MR-U-Net \cite{sahayam2022brain}, proposed by the authors for brain tumor segmentation has also been used to study the proposed workflow. A summary of each of the models is as follows,

\subsubsection{U-Net} 
The U-Net \cite{ronneberger2015u} architecture is a convolutional neural network (CNN) designed for the semantic segmentation of medical images. It consists of contracting paths, expansion  paths, and skip connections. The contraction path has convolution layers followed by max pooling layers. The convolution layers capture the context of the image and the pooling layers down-samples the image. Pooling layers helps to increase the receptive field of the lower layers enabling them to learn neighbor features. The expansion path up-samples and reconstructs the region of interest. It produces the final segmentation mask at the final layer. The skip connection is between contracting and expanding paths at the same level. It helps reconstruct the target image by providing spatial information lost during the contracting path.

\subsubsection{V-Net} The V-Net\cite{milletari2016v} model is an improvement over the U-Net model designed to handle volumetric data for medical image segmentation. It performs 3D convolution to learn spatial and volumetric features unlike 2D convolutions used in U-Net. It also includes residual connections that help to avoid the vanishing gradient problem. It has been successfully applied to various segmentation tasks relating to multiple sclerosis lesion segmentation, and head and neck tumor segmentation.

\subsubsection{Attention U-Net} The Attention U-Net \cite{oktay2018attention} incorporates the attention mechanism into the traditional U-Net model. The attention mechanism learns to assign weights to features based on their relevance to the task. The attention can be applied over a spatial region or across channels. In the attention U-Net, the attention mechanism is integrated at the point where the skip connection from the contraction path in the U-Net has merged with the expansion up-convolution path. The aim is to identify important features when merging high-level features from the encoder and low-level features from the decoder.

\subsubsection{U-Net 3Plus}  The U-Net 3+ \cite{huang2020unet} architecture improves the original U-Net with multi-scale connections, multiple supervisions, and dense connections. The multi-scale connections capture contextual information of various filter sizes. It is useful in dealing with objects of varying sizes like tumors. The skip connections from the encoder to the decoder are dense in such a way that the output of an encoder is concatenated with the decoder at the same level and all other decoders below the current level. The supervision is done at every decoder level and it directly impacts the features learned at the decoder level. Additionally, each decoder from the lower level is concatenated to the immediate higher decoder level and all other decoder levels above.

\subsubsection{Multi
Resolution U-Net with Residual Dual Attention and
Deep Supervision (MRA-U-Net with RDAD)} Hybrid MR-U-Net\cite{sahayam2022brain} is an improved version of the U-Net model proposed in an earlier work by the authors. Each convolution block in the U-Net is replaced with a multi-resolution block that has been used to extract features of varying scales. Skip connections merge the high-level features from the encoder with the low-level features from the decoder which results in a semantic gap. The authors proposed using residual connections and dual attention blocks to handle the mismatch. A deep supervision block concatenates the output from various decoder blocks and reconstructs the final target segmentation. It is expected to generate a better final prediction as it gets feedback from all the decoder blocks rather than just the final layer.
\subsubsection{Swin U-Net} The \cite{cao2023swin} Swin U-Net is a transformer-based architecture with an encoder-decoder structure like a U-Net with patch extraction, patch embedding, patch merging, patch expansion, and swin transformer blocks. The patch extraction block divides the input image into patches of smaller sizes. Features from the patches are extracted by an artificial neural network (ANN) at the patch embedding layer. The patch merging between the swin transformer blocks combines a collection of non-overlapping $n$ patches ($n = 4$) from the higher level into a single patch and increases the channels by $n$ in the lower level. It decreases the number of patches enabling the learning of important features in the swin transformer block. The reduction in the number of patches is done using a linear fully connected neural network. The patch expansion block does the reverse process of the patch merging block to reconstruct the target image. The swin transformer block is named after shifted window self-attention layer. The self-attention patch in transformers needs to compute attention with all other patches and it is computationally expensive. The window self-attention layer groups a collection of non-overlapping neighboring patches into windows. Self-attention is calculated on the patches within a window and the patches in the other window are ignored thereby reducing computation time. Shifted window shifts the entire window by $k$ number of patches after calculating self-attention similar to a normal convolution. The space after shifting is filled by the patches that fall out of the window and the process is called cyclic shifting.

\section{Experimental Results}
\label{chap4}
The experimental section details the dataset used for the study, the implementation details, the performance metrics used to measure the effectiveness of the models and the methodology, tabulating the model output, and presenting the learned edge maps and activation maps for explainability.

\subsection{Dataset}
The Brain Tumor Segmentation Challenge (BraTS) \cite{menze2014multimodal, bakas2018identifying, Bakas2017} is an annual challenge that provides a benchmark dataset for the segmentation of brain tumors. It also provides an online leaderboard and an evaluation engine for analyzing various brain tumor segmentation models. BraTS2020 \cite{bakas2017segmentation, bakas2017segmentationgbm} is a dataset that contains both high-grade gliomas(HGG) and low-grade gliomas. The dataset has 369 patients for training and 125 patients for validation. Each of the patients has four MRI sequences: T1-weighted (T1), T1-weighted contrast-enhanced (T1CE), T2-weighted (T2), and Fluid Attenuated Inversion Recovery (FLAIR). The T1-weighted image generally contains no hyper-intense tumor region and has not been used in the proposed work. The BraTS2020 dataset contains patients from both previous BraTS challenges and it can be considered as a super-set of BraTS2018 and BraTS2019 datasets.

The MR images are preprocessed such that the skulls and neck regions are removed. The images are co-registered and $1$ voxel in the image corresponds to $1 mm^{3}$ of brain tissue. As discussed in \ref{chap1}, the ground truth consists of peritumoral edema (ED) marked in light grey given by an intensity value of $2$,  enhancing tumor (ET) represented as a white region with $4$ as an intensity value and the non-enhancing tumor (NET) and necrotic core region (NCR) as dark grey with an intensity value of $1$. The models are evaluated as enhancing tumor (ET), tumor core (TC), and whole tumor (WT). The tumor core corresponds to NET/NCR and the enhancing tumor region. The whole tumor corresponds to all three tumor regions put together. The goal is to learn ED, NET/NCR, and ET regions.

\subsection{Implementation Details}
The axial view of each of the patients has been used in the training process. None of the frames are dropped. For a patient, a $240$ x $240$ 2D axial slice is stacked in such a way that the input image would be of dimension $240$ x $240$ x $3$. The stacked frames are the corresponding frames from FLAIR, T1CE, and T2 MR images for a patient. A patient will have $155$ such stacked 2D axial input frames. After stacking the input image would be $155$ x $240$ x $240$ x $3$. The Z-score normalization of the inputs, edge extraction from the ground truth images, and the one-hot representation of the ground truth and the edges have been performed as described in Section \ref{chap3}. The one-hot images are stacked on top of each other. If the training consisted of only the tumor regions, the stacked one-hot representation will have a dimension of $155$ x $240$ x $240$ x $4$ with the last dimension corresponding to the background, NET/NCR, edema, and enhancing, respectively with $155$ slices per patient. If the training consisted of tumors and tumor edges as targets, then the dimension would be $155$ x $240$ x $240$ x $7$ with the last dimension consisting of the tumor edges one-hot representation and the tumor one-hot representation. For the U-Net models, the starting number of filters is kept at $32$ and is increased by a factor of $2$ at each depth up to $3$. The bottleneck layer has $256$ filters in total. So, level $1$ has $32$ filters, level $2$ has $64$ filters, level $3$ has $128$ filters, and level $4$ has $256$ filters. All the models used ReLU as an activation function for a fair comparison and SoftMax activation in the last layer to get the final tumor and/or tumor edge prediction. The Swin U-Net consists of 128 neurons in its linear ANN layers to reduce the number of computations. A batch size of $64$ has been used for training all the models. Adam \cite{kingma2014adam} has been used as an optimizer with an initial learning rate $\alpha$ of $0.01$. The models have been trained for a total of $50$ epochs. The model weights are saved for epochs with the highest dice score value.

The experiments are carried out using TensorFlow Keras \cite{chollet2015keras} API. The models have been taken from the Keras U-Net Collections \cite{keras-unet-collection}. The models have been trained on NVIDIA A100 with 40GB RAM.

\subsection{Performance Metrics}
Performance metrics are used to evaluate the effectiveness of the predicted segmentation when compared with the ground truth. The metrics used for measuring the segmentation performance are the dice score and the bi-direction Hausdorff95 distance. 

True positive (TP) is the total number of true class instances correctly predicted as true, false positive (FP) is the total number of false class instances incorrectly predicted as true, and false negative (FN) is the total number of true class instances incorrectly predicted as false. The dice score captures the total number of correct predictions by the model and it is given by the equation \ref{dice}.

\begin{equation}
    \textrm{Dice Score =} \frac{2*TP}{(TP + FP) + (TP + FN)}
    \label{dice}
\end{equation}
Hausdorff distance is a metric used in image segmentation tasks to quantify the difference between two sets of points. The Hausdorff distance is defined as the maximum distance between a point in one set and its nearest point in the other set. It helps to identify whether the predicted segmentation mask and the ground truth segmentation mask are identical across the edge regions \cite{huttenlocher1993comparing}. Higher values of Hausdorff distance indicate a poor model performance between the predicted and ground truth. The general one-sided HD from X to Y and Y to X are given by equations \ref{eq4} and \ref{eq5}.
\begin{equation}
        \textrm{$\hat{H}$(X,Y) = } max(min_{x \in X}d(x,Y))
        \label{eq4}
\end{equation}
\begin{equation}
    \textrm{$\hat{H}$(Y,X) = } max(min_{y \in Y}d(X,y))\
    \label{eq5}
\end{equation}

The one-sided HD distance is not a metric. It fails the symmetry property of a metric. However, the bidirectional HD is a metric. The bidirectional HD between these two sets is defined in equation \ref{eq6}.
\begin{equation}
    \textrm{H(X,Y) = }max(\hat{H}(X,Y), \hat{H}(X,Y))
    \label{eq6}
\end{equation}
Hausdorff distance is sensitive to outliers and noise. So, 95\% of the Hausdorff value is used with a 5\% threshold for error.

The BraTS2020 dataset suffers from a high-class imbalance. Focal loss is a loss function designed to handle a class imbalance in binary classification tasks. Focal loss assigns a low weight to the class in a majority of the instances. The weight is managed by a parameter $\alpha$. $\gamma$ is a modulating factor that reduces the weight of easy examples while increasing the weight of hard examples. A $\gamma$ value of $2$ is set for training all of the different deep learning models in the proposed work.
\begin{equation}
   \textrm{Focal Loss $L(y,\hat{y})$ = }-\alpha y(1-\hat{y})^{\gamma}log(\hat{y})-(1-y)\hat{y}^{\gamma}log(1-\hat{y})
\end{equation}
The value of alpha varies depending on the target class as follows,
 where \[
    \alpha= 
\begin{cases}
0.9,& \text{if }  \textrm{target} \in \{\textrm{NET/NCR Edge, Edema Edge, ET Edge}\}\\
   0.8,& \text{if } \textrm{target} \in \{\textrm{NET/NCR, Edema, ET}\}\\
    0.2,              & \textrm{Background}
\end{cases}
\]

\subsection{Quantitative Evaluation}
Table \ref{tab:time-table} shows different models used for brain tumor segmentation with the number of parameters, training time per patient, and prediction time for a single patient using GPU and CPU. The number of parameters is based on the implementation details discussed in the earlier section. The Swin and the hybrid MR-U-Nets have a reduced number of parameters so that models will fit in memory for a batch size of $64$ and their performance can be compared with the other models. It also can be observed that passing the edge as a target only changes the last layer resulting in a small amount of increase in parameters and it shows that such an implementation can be extended to any semantic segmentation model. The time taken for passing through one patient depends on the complexity of the model with Swin and Hybrid MR-U-Nets taking the most time. From a deployment perspective, all medical institutions may not have GPU support and so, the prediction time for a patient is calculated for both GPU and CPU. It can be observed that the models in general, take around 4 to 5 seconds to predict the tumor regions per patient on a CPU. Overall, the table highlights the trade-off between model complexity and prediction time. It emphasizes the importance of choosing an appropriate model for a given application based on computational resources and time constraints.

\begin{table}[tbh]
\caption{Shows the number parameters for each of the models, the amount of training time taken for one patient in the A100 GPU, and the prediction time per patient in GPU as well as CPU.}
\resizebox{\textwidth}{!}{
\begin{tabular}{ccccc}
\hline
Model                                     & \# of Parameters & \begin{tabular}[c]{@{}c@{}}Training time \\ (sec) per patient\\ (on GPU)\end{tabular} & \begin{tabular}[c]{@{}c@{}}Prediction time \\ (sec) per patient\\ (on GPU)\end{tabular} & \begin{tabular}[c]{@{}c@{}}Prediction time \\ (sec) per patient \\ (on CPU)\end{tabular} \\ \hline
U-Net \cite{ronneberger2015u}             & 2,145,860       & 1.22                                                                                  & 0.58                                                                                    & 3.75                                                                                     \\
U-Net Edge                                & 2,145,959       & 1.23                                                                                  & 0.61                                                                                    & 3.73                                                                                     \\ \hline
V-Net \cite{milletari2016v}               & 3,799,236       & 1.24                                                                                  & 0.62                                                                                    & 4.25                                                                                     \\
V-Net Edge                                & 3,799,335       & 1.25                                                                                  & 0.65                                                                                    & 4.33                                                                                     \\ \hline
Attention U-Net \cite{oktay2018attention} & 2,167,703       & 1.24                                                                                  & 0.65                                                                                    & 3.97                                                                                     \\
Attention U-Net Edge                      & 2,167,802       & 1.24                                                                                  & 0.62                                                                                    & 4.00                                                                                     \\ \hline
U-Net 3 Plus \cite{huang2020unet}         & 1,994,860       & 1.44                                                                                  & 0.78                                                                                    & 5.32                                                                                     \\
U-Net 3 Plus Edge                         & 2,013,052       & 2.46                                                                                  & 0.92                                                                                    & 5.41                                                                                     \\ \hline
Swin U-Net  \cite{cao2023swin}            & 715,532         & 2.56                                                                                  & 0.81                                                                                    & 4.49                                                                                     \\
Swin U-Net Edge                           & 715,559         & 2.55                                                                                  & 0.89                                                                                    & 4.54                                                                                     \\ \hline
Hybrid MR-U-Net \cite{sahayam2022brain}       & 505,559         & 2.53                                                                                  & 0.65                                                                                    & 3.43                                                                                     \\
Hybrid MR-U-Net Edge                         & 505,598         & 2.54                                                                                  & 0.69                                                                                    & 3.45                                                                                     \\ \hline
\end{tabular}
}
\label{tab:time-table}
\end{table}
Table \ref{tab:mean_training} shows the mean results obtained for the various baseline models and their edge target counterparts. It can be observed the earlier models like U-Net, V-Net, and Attention U-Net show improvements when training with edges as targets in the enhancing and core tumor dice and Hausdorff95 score. All the models except the swin transformer performed well in dice and Hausdorff95 for both tumor core and enhancing tumor regions. The swin transformer did not improve in any of the metrics when learning along with target edges. Additionally, it can be noted that the swin transformer is the only model that uses linear artificial neural networks to learn features instead of convolutional layers. Also, the whole tumor degrades in performance in dice score and Hausdorff95 distance for all the models except U-Net and V-Net in the Hausdorff95 metric. The authors conjecture that the recent models might have blocks that take care of learning edges, like the swin transformer blocks and self-attention blocks in the swin transformer, and the dual attention blocks in hybrid MR-U-Net. However, learning edges along with tumors shows that even earlier models like U-Net and V-Net could give performance close to recent models like swin U-Net and hybrid MR-U-Net.

\begin{table}[H]
\centering
\caption{Comparison of the mean of different models on the training dataset. The values in \textbf{bold} represent the best performance of the existing and proposed models. The values in {\color[HTML]{FE0000} \textbf{red}} are the overall best. \up{} and \invupp{} denotes improvement, and \down{} and \invdownp{} denotes deterioration of the proposed edge method with respect to the normal method.}
\resizebox{\textwidth}{!}{
\begin{tabular}{cllllll}
\hline
                                         & \multicolumn{3}{c}{Dice Score  $\uparrow$}                                                                                                                                               & \multicolumn{3}{c}{Hausdorff95 $\downarrow$}                                                                                                                 \\ \cline{2-7} 
\multirow{-2}{*}{Model}                  & \multicolumn{1}{c}{ET}                                                          & \multicolumn{1}{c}{WT}                                                         & \multicolumn{1}{c}{TC}                                                          & \multicolumn{1}{c}{ET}                                                          & \multicolumn{1}{c}{WT}                                     & \multicolumn{1}{c}{TC}                                                 \\ \hline
U-Net \cite{ronneberger2015u}                   & \textbf{0.747}                     & \textbf{0.898}                         & 0.761                                   &  \textbf{29.80}  & 11.38                                 & 16.79                                                                                              \\

U-Net Edge               & 0.741 \downp{0.80}                                 & 0.894 \downp{0.4}                                 & \textbf{0.775}\upp{1.8}                        & 30.33 \invdownp{1.76}                             & \textbf{07.89}\invupp{30.67}                        & \textbf{11.51}\invupp{31.44}                                                                                      \\ \hline

V-Net \cite{milletari2016v} & 0.749                                  & \textbf{0.900}                        & 0.792                                 & 30.43                               & 9.17                                 & 11.74                                                                     \\

V-Net Edge               &  \textbf{0.755}\upp{0.80} & 0.895 \downp{0.55}                                 & \textbf{0.811}\upp{2.40}                        & \textbf{26.90}\invupp{11.60}                        & \textbf{08.72}\invupp{4.91}                        & \textbf{08.71}\invupp{25.81}                                                             \\ \hline

Attention U-Net \cite{oktay2018attention}          & 0.724                                 & \textbf{0.888}                        & \textbf{0.782}                        & 34.82                                 & \textbf{13.46}                        & 13.10                                                           \\

Attention U-Net Edge     & \textbf{0.732}\upp{1.10}                        & 0.848 \downp{4.50}                                 & 0.767 \downp{1.92}                                 & \textbf{32.19}\invupp{7.55}                         & 29.79 \invdownp{122.32}                                & \textbf{10.67}\invupp{18.55}                                                            \\ \hline

U-Net 3 Plus \cite{huang2020unet}            & 0.725                                 & \textbf{0.897}                        & 0.780                                 & 35.54                                  & \textbf{08.72}                        & 13.76                                                                     \\

U-Net 3 Plus Edge        & \textbf{0.738}\upp{1.79}                        & 0.885 \downp{1.34}                                 & \textbf{0.817}\upp{4.74}                         & \textbf{32.05}\invupp{9.81}                        & 11.31 \invdownp{29.70}                                & \textbf{10.74}\invupp{21.95}                                                            \\ \hline

Swin U-Net \cite{cao2023swin}              & 0.743                                 & \color[HTML]{FE0000} \textbf{0.918} & \color[HTML]{FE0000} \textbf{0.871} & \textbf{28.78}                                 & \color[HTML]{FE0000} \textbf{02.96} & \color[HTML]{FE0000} \textbf{4.37} \\

Swin U-Net Edge          & \textbf{0.743}\upp{0.00}                        & 0.913 \downp{0.54}                                 & 0.862 \downp{1.03}                                  & 31.84\invdownp{10.63}                        & 04.40 \invdownp{48.64}                                 & 5.27 \invdownp{20.59}                                                                      \\ \hline

Hybrid MR-U-Net \cite{sahayam2022brain}              & \color[HTML]{FE0000} \textbf{0.765}                                 & \textbf{0.906} & 0.775 & \color[HTML]{FE0000} \textbf{26.19}                                 & \textbf{08.07} &  11.07 \\

Hybrid MR-U-Net Edge          & 0.762 \downp{0.39}                        & 0.882 \downp{2.68}                                 & \textbf{0.824} \upp{6.13}                                  & 30.84\invdownp{16.31}                        & 13.22 \invdownp{48.38}                                 & \textbf{5.71}\invupp{20.59}                                                                      \\ \hline
\end{tabular}
}
\label{tab:mean_training}
\end{table}

Some patients in the dataset do not have an enhancing tumor segmentation region. If a model happens to predict even a single voxel of enhancing tumor for such patients, the BraTS online evaluation engine assigns a maximum penalty of $0$ for dice score and $373.12866$ for Hausdorff95 distance. It can skew the results even if most of the patient's predictions are correct. For the mentioned reason, the authors compared the median dice score and the Hausdorff95 in addition to the mean metric results. The median metric results for the training data are shown in Table \ref{tab:median_training}. A similar observation from the mean training results can be made in the median training results. Another observation from mean and median training results is that the models with training along with edges performed well for smaller tumor regions like the enhancing tumor region and tumor core region.

\begin{table}[H]
\centering
\caption{Comparison of the median of different models on the training dataset. The values in \textbf{bold} represent the best performance of the existing and proposed models. The values in {\color[HTML]{FE0000} \textbf{red}} are the overall best. \up{} and \invupp{} denotes improvement, and \down{} and \invdownp{} denotes deterioration of the proposed edge method with respect to the normal method.}
\resizebox{\textwidth}{!}{
\begin{tabular}{cllllll}
\hline
                                         & \multicolumn{3}{c}{Dice Score  $\uparrow$}                                                                                                                                               & \multicolumn{3}{c}{Hausdorff95 $\downarrow$}                                                                                                                 \\ \cline{2-7} 
\multirow{-2}{*}{Model}                  & \multicolumn{1}{c}{ET}                                                          & \multicolumn{1}{c}{WT}                                                         & \multicolumn{1}{c}{TC}                                                          & \multicolumn{1}{c}{ET}                                                          & \multicolumn{1}{c}{WT}                                     & \multicolumn{1}{c}{TC}                                                 \\ \hline
U-Net \cite{ronneberger2015u}             & \color[HTML]{000000} 0.850          & \color[HTML]{000000} \textbf{0.923} & \color[HTML]{000000} 0.887          & \color[HTML]{000000} \textbf{2.00} & \color[HTML]{000000} 3.74          & \color[HTML]{000000} 3.61                                                                \\
U-Net Edge                                & \color[HTML]{000000} \textbf{0.858}\upp{0.94} & \color[HTML]{000000} 0.919 \downp{0.43}         & \color[HTML]{000000} \textbf{0.909}\upp{2.48} & \color[HTML]{000000} \textbf{2.00}\invupp{0.00} & \color[HTML]{000000} \textbf{3.61}\invupp{3.47} & \color[HTML]{000000} \textbf{3.00}\invupp{16.89}                                                              \\ \hline
V-Net \cite{milletari2016v}               & \color[HTML]{000000} \textbf{0.858}         & \color[HTML]{000000} \textbf{0.923} & \color[HTML]{000000} 0.906          & \color[HTML]{000000} 2.00          & \color[HTML]{000000} \textbf{3.46}  & \color[HTML]{000000} 3.605                                                                 \\
V-Net Edge                                & { \textbf{0.858}}\upp{0.00}                    & {\color[HTML]{000000} 0.920} \downp{0.32}                           & {\color[HTML]{000000} \textbf{0.913}}  \upp{0.77}                  & {\color[HTML]{000000} \textbf{1.73}}    \invupp{13.50}          & {\color[HTML]{000000} 4.00}   \invdownp{16.61}                                 & {\color[HTML]{FE0000} \textbf{2.82}}   \invupp{21.67}                                                                  \\ \hline
Attention U-Net \cite{oktay2018attention} & 0.841                               & \textbf{0.912}                     & \color[HTML]{000000} 0.891                            & \color[HTML]{000000} \textbf{2.23} & \color[HTML]{000000} \textbf{4.69}                     & \color[HTML]{000000} 4.24                                                                               \\
Attention U-Net Edge                      & {\color[HTML]{000000} \textbf{0.851}}\upp{1.189}                    & {\color[HTML]{000000} 0.887} \downp{2.74}                           & {\color[HTML]{000000} \textbf{0.897}} \upp{0.67}                   & {\color[HTML]{000000} \textbf{2.24}}  \invdownp{0.45}             & {\color[HTML]{000000} 22.17}   \invdownp{372.707}                           & {\color[HTML]{000000} \textbf{4.13}}   \invupp{2.59}                                                                \\ \hline
U-Net 3 Plus \cite{huang2020unet}         & \color[HTML]{000000} \textbf{0.846}                     & \color[HTML]{000000} \textbf{0.922}                     & \color[HTML]{000000} \textbf{0.905}                     & \color[HTML]{000000} \textbf{2.23}               & \color[HTML]{000000} \textbf{3.31}                     & \color[HTML]{000000} \textbf{3.16}                                                                      \\
U-Net 3 Plus Edge                         & {\color[HTML]{000000} 0.843} \downp{0.35}                             & {\color[HTML]{000000} 0.907} \downp{1.63}                             & {\color[HTML]{000000} 0.896} \downp{0.995}                           & {\color[HTML]{000000} \textbf{2.23}} \invupp{0.00}             & {\color[HTML]{000000} 4.24} \invdownp{28.10}                            & {\color[HTML]{000000} 4.24}        \invdownp{34.177}                                                                      \\ \hline

Swin U-Net  \cite{cao2023swin}            & \color[HTML]{000000} 0.851                               & \color[HTML]{FE0000} \textbf{0.933}                     & \color[HTML]{000000} 0.909                              & \color[HTML]{000000} \textbf{1.73}               & \color[HTML]{FE0000} \textbf{2.23}                     & \color[HTML]{000000} \textbf{3.00}                                                                           \\
Swin U-Net Edge                           & {\color[HTML]{000000} \textbf{0.854}} \upp{0.35}                   & {\color[HTML]{000000} 0.928} \downp{0.54}                            & {\color[HTML]{000000} \textbf{0.909}}\upp{0.00}                    & {\color[HTML]{000000} 2.00}  \invdownp{15.60}                            & {\color[HTML]{000000} 2.44} \invdownp{4.72}                             & {\color[HTML]{000000} 3.16}   \invdownp{5.33}                                                                           \\ \hline

Hybrid MR-U-Net  \cite{sahayam2022brain}            & \color[HTML]{000000} 0.866                               & \textbf{0.927}                     & \color[HTML]{000000} 0.911                              &  1.73               &  \textbf{3.00}                     & \color[HTML]{000000} \textbf{2.82}                                                                           \\
Hybrid MR-U-Net Edge                           & {\color[HTML]{FE0000} \textbf{0.867}} \upp{0.11}                   & {\color[HTML]{000000} 0.908} \downp{2.05}                            & {\color[HTML]{FE0000} \textbf{0.915}}\upp{0.439}                    & {\color[HTML]{FE0000} \textbf{1.41}}  \invupp{18.50}                            & {\color[HTML]{000000} 5.10} \invdownp{70.00}                             & {\color[HTML]{000000} 3.00}   \invdownp{6.38}                                                                           \\ \hline
\end{tabular}
}
\label{tab:median_training}
\end{table}

Tables \ref{tab:mean_validation} and \ref{tab:median_validation} show the mean, and median dice and Hausdorff95 results on the validation dataset. The difference between the models trained with and without edges shows similar observations as in the training data. However, from the training data, the dice score for enhancing and tumor core has fallen significantly. It  does well for the whole tumor region but not for enhancing, and core tumor regions. The fall in the performance of enhancing tumor and tumor core is observed mostly in patients that do not have an enhancing tumor region resulting in a maximum penalty from the evaluation engine. The improved enhancing and tumor core performance in the median validation results show the results affected in the mean validation results due to the maximum penalty.

\begin{table}[H]
\caption{Comparison of the mean of different models on the validation dataset. The values in \textbf{bold} represent the best performance of the existing and proposed models. The values in {\color[HTML]{FE0000} \textbf{red}} are the overall best. \up{} and \invupp{} denotes improvement, and \down{} and \invdownp{} denotes deterioration of the proposed edge method with respect to the normal method.}
\resizebox{\textwidth}{!}{
\begin{tabular}{cllllll}
\hline
                                         & \multicolumn{3}{c}{Dice Score  $\uparrow$}                                                                                                                                               & \multicolumn{3}{c}{Hausdorff95 $\downarrow$}                                                                                                                 \\ \cline{2-7} 
\multirow{-2}{*}{Model}                  & \multicolumn{1}{c}{ET}                                                          & \multicolumn{1}{c}{WT}                                                         & \multicolumn{1}{c}{TC}                                                          & \multicolumn{1}{c}{ET}                                                          & \multicolumn{1}{c}{WT}                                     & \multicolumn{1}{c}{TC}                                                 \\ \hline
U-Net \cite{ronneberger2015u}             & {\color[HTML]{000000} 0.688}          & \color[HTML]{000000} \textbf{0.875} & \color[HTML]{000000} 0.649          & \color[HTML]{000000} \textbf{36.15} & \color[HTML]{000000} 15.15          & \color[HTML]{000000} 26.90       \\
U-Net Edge                                & \color[HTML]{FE0000} \textbf{0.692} \color[HTML]{000000} \upp{0.58} & \color[HTML]{000000} 0.867 \downp{0.91}          & \color[HTML]{000000} \textbf{0.671} \upp{3.38} & \color[HTML]{000000} 38.69 \invdownp{7.02}          & \color[HTML]{000000} \textbf{13.41} \invupp{11.48} & \color[HTML]{000000} \textbf{22.83} \invupp{15.13}                                                                            \\ \hline
V-Net \cite{milletari2016v}               & \color[HTML]{000000} 0.664           & \color[HTML]{000000} \textbf{0.873} & \color[HTML]{000000} 0.673          & \color[HTML]{000000} 51.19          & \color[HTML]{000000} 15.94          & \color[HTML]{000000} 24.69                                                                                    \\
V-Net Edge                                & {\color[HTML]{000000} \textbf{0.678}} \upp{2.11}                    & {\color[HTML]{000000} 0.869} \downp{0.46}                              & {\color[HTML]{000000} \textbf{0.675}}  \upp{0.30}                   & {\color[HTML]{000000} \textbf{42.39}}   \invupp{17.19}                  & {\color[HTML]{000000} \textbf{13.66}} \invupp{14.30}                      & {\color[HTML]{000000} \textbf{21.14}}  \invupp{14.37}                                                                                  \\ \hline
Attention U-Net \cite{oktay2018attention} & {\color[HTML]{000000} 0.646}                              & {\color[HTML]{000000} \textbf{0.863}}                     & {\color[HTML]{000000} \textbf{0.679}}                     & {\color[HTML]{000000} 56.37}                              & {\color[HTML]{000000} \textbf{16.81}}                     & {\color[HTML]{000000} 27.09}                \\
Attention U-Net Edge                      & {\color[HTML]{000000} \textbf{0.660}}  \upp{2.17}                   & {\color[HTML]{000000} 0.820} \downp{4.98}                             & {\color[HTML]{000000} 0.656}  \downp{3.38}                            & {\color[HTML]{000000} \textbf{49.53}}\invupp{12.13}                    & {\color[HTML]{000000} 35.18} \invdownp{109.28}                              & {\color[HTML]{000000} \textbf{17.36}} \invupp{56.04}                                                                                   \\ \hline
U-Net 3 Plus \cite{huang2020unet}         & {\color[HTML]{000000} \textbf{0.663}}                     & {\color[HTML]{000000} \textbf{0.868}}                     & {\color[HTML]{000000} 0.665}                              & {\color[HTML]{000000} \textbf{48.21}}                     & {\color[HTML]{000000} \textbf{14.50}}                              & {\color[HTML]{000000} 26.47}                \\
U-Net 3 Plus Edge                         & {\color[HTML]{000000} 0.663}  \downp{0.00}                            & {\color[HTML]{000000} 0.862}  \downp{0.69}                             & {\color[HTML]{FE0000} \textbf{0.693}} \upp{4.21}                   & {\color[HTML]{000000} 49.44}  \invdownp{2.55}                           & {\color[HTML]{000000} 14.81} \invdownp{2.13}                    & {\color[HTML]{000000} \textbf{21.13}} \invupp{20.17}                                                                                  \\ \hline
Swin U-Net  \cite{cao2023swin}            & {\color[HTML]{000000} \textbf{0.641}}                     & {\color[HTML]{000000} \textbf{0.860}}                     & {\color[HTML]{000000} \textbf{0.682}}                     & {\color[HTML]{000000} \textbf{43.57}}                     & {\color[HTML]{FE0000} \textbf{11.13}}                     & {\color[HTML]{FE0000} \textbf{16.95}}                                                                            \\
Swin U-Net Edge                           & {\color[HTML]{000000} 0.628} \downp{2.02}                         & {\color[HTML]{000000} 0.858} \downp{0.23}                        & {\color[HTML]{000000} 0.675} \downp{1.02}                       & {\color[HTML]{000000} 48.48} \invdownp{11.26}                       & {\color[HTML]{000000} 13.81} \invdownp{24.07}         & {\color[HTML]{000000} 21.31} \invdownp{25.72}                                                                                         \\ \hline

Hybrid MR-U-Net  \cite{sahayam2022brain}            & \color[HTML]{000000} \textbf{0.691}                              & \color[HTML]{FE0000}\textbf{0.876}                     & \color[HTML]{000000} 0.644                              &  \color[HTML]{FE0000} \textbf{34.43}               &  \textbf{14.58}                     & \color[HTML]{000000} 32.93                                                                           \\
Hybrid MR-U-Net Edge                           & {\color[HTML]{000000} 0.681} \downp{1.44}                   & {\color[HTML]{000000} 0.847} \downp{3.31}                            & {\color[HTML]{FE0000} \textbf{0.693}}\upp{7.60}                    & {40.70}  \invdownp{18.21}                            & {\color[HTML]{000000} 19.76} \invdownp{35.52}                             & {\color[HTML]{000000} \textbf{20.88}}   \invupp{36.59}                                                                           \\ \hline

\end{tabular}
}
\label{tab:mean_validation}
\end{table}

\begin{table}[H]
\caption{Comparison of the median of different models on the validation dataset. The values in \textbf{bold} represent the best performance of the existing and proposed models. The values in {\color[HTML]{FE0000} \textbf{red}} are the overall best. \up{} and \invupp{} denotes improvement, and \down{} and \invdownp{} denotes deterioration of the proposed edge method with respect to the normal method.}
\resizebox{\textwidth}{!}{
\begin{tabular}{cllllll}
\hline
                                         & \multicolumn{3}{c}{Dice Score  $\uparrow$}                                                                                                                                               & \multicolumn{3}{c}{Hausdorff95 $\downarrow$}                                                                                                                 \\ \cline{2-7} 
\multirow{-2}{*}{Model}                  & \multicolumn{1}{c}{ET}                                                          & \multicolumn{1}{c}{WT}                                                         & \multicolumn{1}{c}{TC}                                                          & \multicolumn{1}{c}{ET}                                                          & \multicolumn{1}{c}{WT}                                     & \multicolumn{1}{c}{TC}                                                 \\ \hline

U-Net \cite{ronneberger2015u}             & \color[HTML]{FE0000} \textbf{0.837} & \color[HTML]{FE0000} \textbf{0.905} & 0.812                                 & 2.45                                 & \textbf{4.90}   & 5.75                                                                                    \\

U-Net Edge                                & 0.835 \downp{0.23}                                  & 0.901 \downp{0.44}                                & {\color[HTML]{FE0000} \textbf{0.867}} \upp{6.77} & {\color[HTML]{FE0000} \textbf{2.35}} \invupp{4.08} & 5.00 \invdownp{2.04}                  & {\color[HTML]{FE0000} \textbf{4.243}} \invupp{26.43}                                                                          \\ \hline

V-Net \cite{milletari2016v}               & 0.811                                 & \textbf{0.899}                       & 0.841                                 & 3.61                                 & 5.75            & 7.81                        
\\

V-Net Edge                                & \textbf{0.830} \upp{2.34}                                            & 0.894 \downp{0.55}                                                    & \textbf{0.860} \upp{2.25}                                            & \textbf{2.45} \invupp{32.13}                                            & \textbf{5.47} \invupp{4.86}                       & \textbf{5.09} \invupp{34.82}                                                                                                             \\ \hline

Attention U-Net \cite{oktay2018attention} & \textbf{0.808}                                            & \textbf{0.901}                                           & 0.838                                                     & 4.582                                                     & \textbf{6.40}                       & 9.11                                                                                                                       \\

Attention U-Net Edge                      & 0.81 \downp{0.24}                                                    & 0.86 \downp{4.55}                                                    & \textbf{0.845} \upp{0.83}                                           & \textbf{3.61} \invupp{21.21}                                            & 29.581 \invdownp{53.90}                                & \textbf{6.16} \invupp{32.38}                                                                          \\ \hline

U-Net 3 Plus \cite{huang2020unet}         & \textbf{0.825}                                            & \textbf{0.903}                                           & \textbf{0.840}                                            & 3.74                                                     & \textbf{4.90}                       & \textbf{7.07}                                                                            \\

U-Net 3 Plus Edge                         & 0.815 \downp{1.21}                                                     & 0.890 \downp{1.43}                                                     & 0.830 \downp{1.19}                                                     & \textbf{3.46} \invupp{7.48}                                            & 5.47 \invdownp{11.63}                                & 8.77 \invdownp{24.04}                                                                         \\ \hline

Swin U-Net  \cite{cao2023swin}            & \textbf{0.804}                                            & \textbf{0.903}                                           & 0.821                                                     & \textbf{3.16}                                            & {\color[HTML]{FE0000} \textbf{4.35}} & \textbf{7.141}                                                                                             \\

Swin U-Net Edge                           & 0.804 \upp{0.00}                                                     & 0.894 \downp{0.99}                                                   & \textbf{0.827} \upp{0.73}                                           & 3.16 \invupp{0.00}                                                    & 5.00 \invdownp{14.94}                                      & 7.68 \invdownp{7.56} 
\\ \hline

Hybrid MR-U-Net  \cite{sahayam2022brain}            & \color[HTML]{000000} \textbf{0.827}                              & \color[HTML]{000000}\textbf{0.899}                     & \color[HTML]{000000} 0.841                              &  2.83             &  \textbf{5.48}                     & \color[HTML]{000000} 5.10                                                                          \\
Hybrid MR-U-Net Edge                           & {\color[HTML]{000000} 0.826} \downp{0.12}                   & {\color[HTML]{000000} 0.880} \downp{2.11}                            & {\color[HTML]{000000} \textbf{0.857}}\upp{1.90}                    & {\textbf{2.45}}  \invupp{13.42}                            & {\color[HTML]{000000} 8.06} \invdownp{47.08}                             & {\color[HTML]{000000} \textbf{5.00}}   \invupp{1.96}                                                                           \\ \hline
\end{tabular}
}
\label{tab:median_validation}
\end{table}

\subsection{Edge Maps}
\begin{figure}[bht]
\captionsetup[subfigure]{justification=centering}
	\centering
	\subfloat[Prediction along with the Edges]{{\includegraphics[width=0.2\linewidth,height=0.2\linewidth]{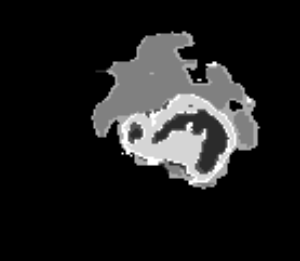} }}
        \quad  
        \subfloat[Highlighted edema edge prediction]{{\includegraphics[width=0.2\linewidth,height=0.2\linewidth]{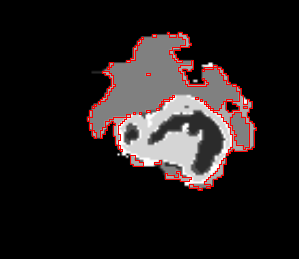} }}
        \quad
	\subfloat[Highlighted NCT/NET edge prediction]{{\includegraphics[width=0.2\linewidth,height=0.2\linewidth]{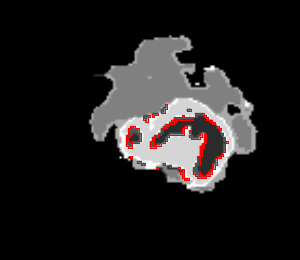} }}
	\quad
     \subfloat[Highlighted ET edge prediction]{{\includegraphics[width=0.2\linewidth,height=0.2\linewidth]{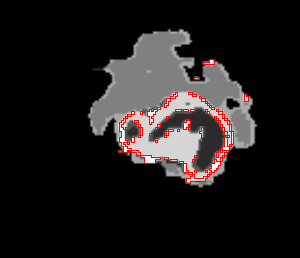} }}
	\caption{Shows the sample set of 2D MR images with predicted tumor edges and tumor regions. The tumor edge regions are highlighted with a red boundary.}

\label{output}
\end{figure}

All the models that learn edges along with the tumor regions will generate an edge map. To show that the models learn edges along with the tumor region, a 2D axial slice of an edge map along with the tumor region is predicted for the V-Net model and it is shown in Figure \ref{output}. Figure \ref{output} (a) shows a sample 2D axial slice prediction without highlighting any of the edges. Figure \ref{output} (b) - (c) shows the 2D axial slice prediction of the edges highlighted with a red boundary for edema, NCT/NET, and ET, respectively. Figure \ref{output} shows that the model predicts the circumference of the tumor region as edges and inner regions as tumor regions. This information maybe be useful for treatment planning when there is a need for getting the edge regions of the tumor areas.

\subsection{Activation Maps}
\begin{figure}[H]
\captionsetup[subfigure]{justification=centering}
	\centering
	\subfloat[Background One-Hot 2D MR Slice]{\includegraphics[width=0.2\linewidth,height=0.2\linewidth]{4background.png}}
        \quad  
        \subfloat[Edema One-Hot 2D MR Slice]{{\includegraphics[width=0.2\linewidth,height=0.2\linewidth]{4edema.png} }}
        \quad
	\subfloat[NCT/NET One-Hot 2D MR Slice]{{\includegraphics[width=0.2\linewidth,height=0.2\linewidth]{4core.png} }}
	\quad
     \subfloat[Enhancing One-Hot 2D MR Slice]{{\includegraphics[width=0.2\linewidth,height=0.2\linewidth]{4enhancing.png} }}

        \subfloat[Background Activation Map]{{\includegraphics[width=0.22\linewidth,height=0.2\linewidth]{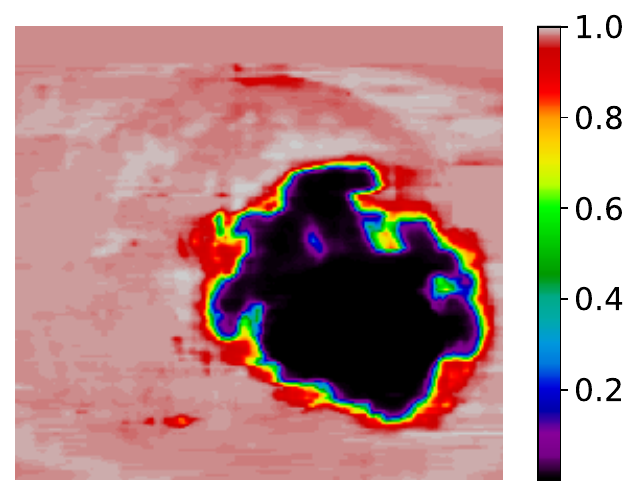} }}
 \subfloat[Edema Activation Map]{{\includegraphics[width=0.22\linewidth,height=0.2\linewidth]{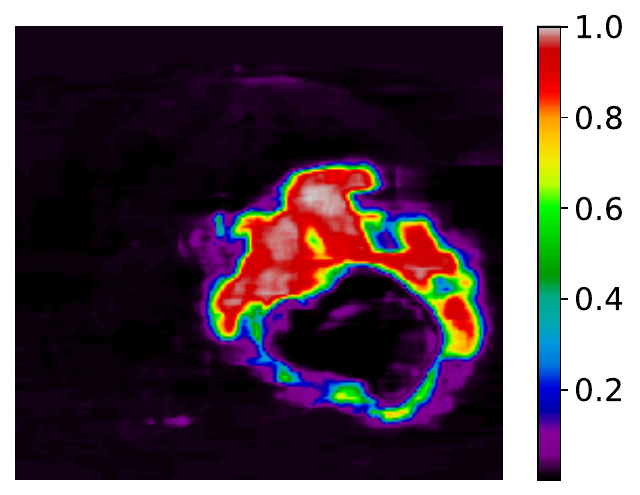} }}
     \subfloat[NCT/NET Activation Map]{{\includegraphics[width=0.22\linewidth,height=0.2\linewidth]{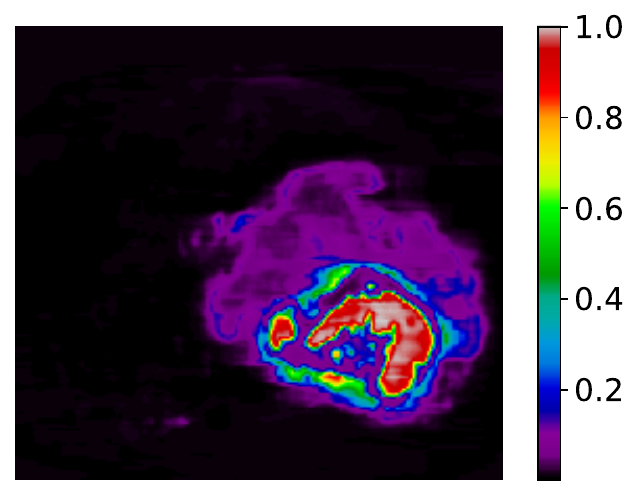} }}
     \subfloat[Enhancing Activation Map]{{\includegraphics[width=0.22\linewidth,height=0.2\linewidth]{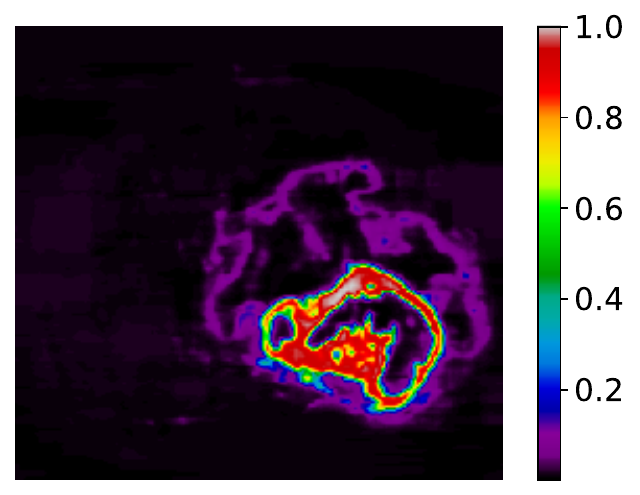} }}

     \subfloat[Background Activation Map]{{\includegraphics[width=0.22\linewidth,height=0.2\linewidth]{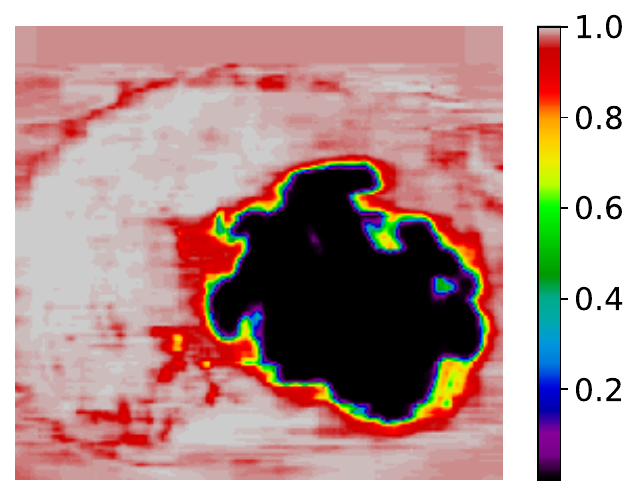} }}
     \subfloat[Edema \\ Activation Map]{{\includegraphics[width=0.22\linewidth,height=0.2\linewidth]{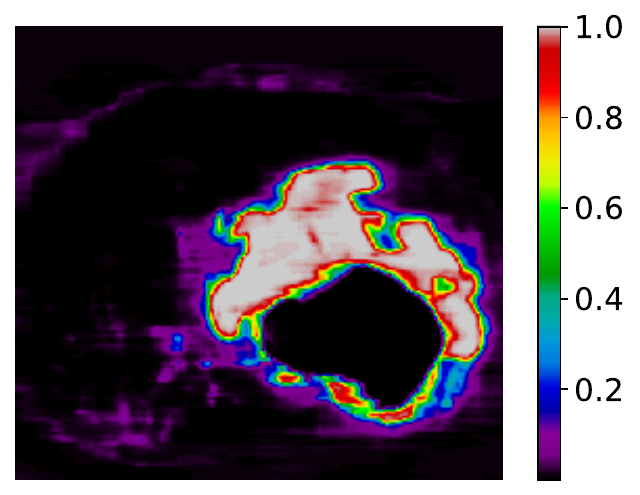} }}
     \subfloat[NCT/NET Activation Map]{{\includegraphics[width=0.22\linewidth,height=0.2\linewidth]{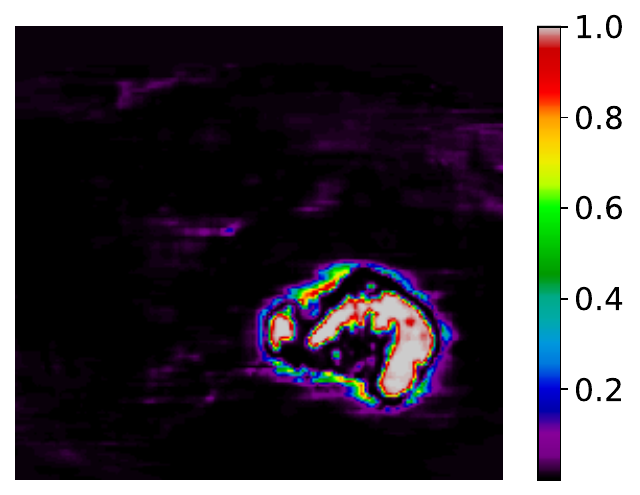} }}
     \subfloat[Enhancing Activation Map]{{\includegraphics[width=0.22\linewidth,height=0.2\linewidth]{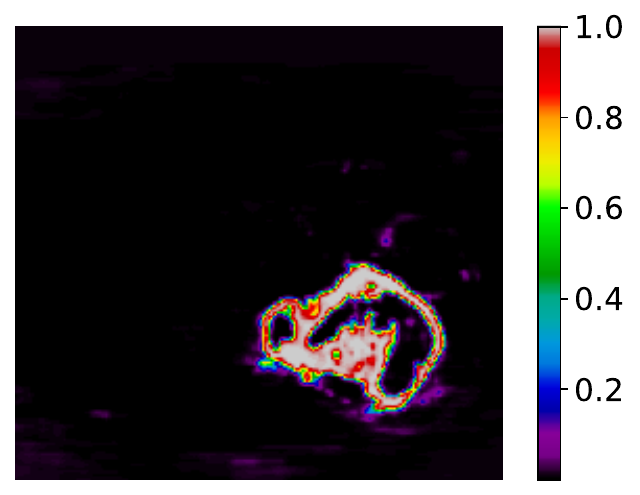} }}

     \caption{(a) to (d) Shows the ground truth one-hot vector for background, edema, NCT/NET, and enhancing tumor. (e) to (h) shows the respective activation maps generated by the hybrid MR-U-Net trained with the tumor regions only. (i) to (l) show the respective activation maps generated by the hybrid MR-U-Net trained with the tumor regions along with the edge regions.}
\label{activationmaps}
\end{figure}

For better explainability of the models trained on tumor regions with and without tumor edges, the authors generated the activation maps by obtaining the predictions from the last layer of each of the models. For example, a sample 2D axial one-hot slice and activation map for hybrid MR-U-Net trained on tumor regions with and without edges is shown in Figure \ref{activationmaps}. The models trained along with the edges appear to have sharper edge activation than models trained without edges as targets. From Figure \ref{activationmaps} (i) to (l), it can be observed that the model trained with tumor regions along with edges had higher activation (closer to $1.0$)  for tumor areas  compared to the model trained with tumor regions only as in Figure \ref{activationmaps} (e)  to (h). Also, the latter has low activation, marked by purple color,  for tumor regions that it shouldn't predict.  For example, the activation of NCT/NET from Figure \ref{activationmaps} (g) shouldn't show activation for edema and enhancing tumor regions, however, the model still has a low activation marked by purple color. A similar activation map is observed for all other models trained for tumor regions with and without edges.

From all the observations from the activation maps, the authors would like to conjecture that the edges of targets might act as data-level attention to learn better tumor and edge segments. Additionally, the higher activation for models trained with tumor regions along with edge regions could be the reason for the improvement in the Hausdorff95 score.

\section{Discussion and Conclusion}
\label{chap6}
In this work, the authors proposed a workflow to learn the edges along with the tumor regions. The aim is to give attention to the tumor edges in addition to the tumor regions. The authors extracted the tumor edges from the ground truth using a derivative-like filter. The final edges have been extracted by utilizing the ground truth as a reference image. Various popular encoder-decoder models like U-Net, V-Net, Attention U-Net, U-Net 3 Plus, Swin U-Net, and Hybrid MR-U-Net have been studied with and without edges as targets along with the tumor regions. The mean results are shown for both the training and validation datasets in Tables \ref{tab:mean_training}, and \ref{tab:mean_validation}. Additionally, the median training and validation results are shown in Tables \ref{tab:median_training}, and \ref{tab:median_validation} due to the high penalty for false positives in the complete absence of a specific tumor region in a patient. The high penalty will skew the mean results and so the median results are discussed to give a clearer picture of the model performance. An edge map is generated for models to show the learned edges for each of the tumor regions. Activation map is studied for models trained on tumor regions with and without edges as targets. From the results obtained, the following observations and conclusions have been made,

\begin{enumerate}
    \item The workflow used for training can be incorporated into any semantic segmentation models with only changes done to the last layer of the architecture. So, the models that learn edges along with the tumor regions will only increase the number of parameters by a small margin. 

    \item  All models in the training and validation dataset, except the swin transformer, show improvements when training with edges as targets in the enhancing and core tumor in the dice and Hausdorff95 score metrics. It shows that the models can learn the smaller tumor regions better when trained with both the tumor regions and edges.
    
    \item Adding edges to the target improved the performance of the U-Net and V-Net models, bringing the edge-trained models closer to the performance of complex models like the swin transformer and hybrid MR-U-Net. It shows that the earlier models may lack the ability to accurately learn the tumor regions in the absence of edge information which demonstrates the usefulness of the proposed training workflow.

    \item The edge map shown in Figure \ref{output} can be used to track the circumference of the tumor regions. It can be useful for treatment and surgery planning.

    \item Activation map from Figure \ref{activationmaps} shows higher activation and more accurate activation for models trained with both tumor regions and tumor edges. The authors conjecture that the edges of targets might act as data-level attention to learn better tumor and edge segments.
\end{enumerate}

As part of future work, the authors aim to propose solutions focusing on improving tumor core and enhancing tumor metrics in the validation phase. Also, the authors aim to handle false positives in cases where the tumor is in a patient.
\bibliographystyle{elsarticle-num}
\bibliography{reference}

\begin{thebibliography}{10}
\expandafter\ifx\csname url\endcsname\relax
  \def\url#1{\texttt{#1}}\fi
\expandafter\ifx\csname urlprefix\endcsname\relax\def\urlprefix{URL }\fi
\expandafter\ifx\csname href\endcsname\relax
  \def\href#1#2{#2} \def\path#1{#1}\fi

\bibitem{ostrom2022cbtrus}
Q.~T. Ostrom, M.~Price, C.~Neff, G.~Cioffi, K.~A. Waite, C.~Kruchko, J.~S.
  Barnholtz-Sloan, Cbtrus statistical report: Primary brain and other central
  nervous system tumors diagnosed in the united states in 2015--2019,
  Neuro-oncology 24~(Supplement\_5) (2022) v1--v95.

\bibitem{mo2020multimodal}
S.~Mo, M.~Cai, L.~Lin, R.~Tong, Q.~Chen, F.~Wang, H.~Hu, Y.~Iwamoto, X.-H. Han,
  Y.-W. Chen, Multimodal priors guided segmentation of liver lesions in mri
  using mutual information based graph co-attention networks, in: Medical Image
  Computing and Computer Assisted Intervention--MICCAI 2020: 23rd International
  Conference, Lima, Peru, October 4--8, 2020, Proceedings, Part IV 23,
  Springer, 2020, pp. 429--438.

\bibitem{sun2019drrnet}
J.~Sun, W.~Chen, S.~Peng, B.~Liu, Drrnet: dense residual refine networks for
  automatic brain tumor segmentation, Journal of Medical Systems 43 (2019)
  1--9.

\bibitem{menze2014multimodal}
B.~H. Menze, A.~Jakab, S.~Bauer, J.~Kalpathy-Cramer, K.~Farahani, J.~Kirby,
  Y.~Burren, N.~Porz, J.~Slotboom, R.~Wiest, et~al., The multimodal brain tumor
  image segmentation benchmark (brats), IEEE Transactions on Medical Imaging
  34~(10) (2014) 1993--2024.

\bibitem{agravat20213d}
R.~R. Agravat, M.~S. Raval, 3d semantic segmentation of brain tumor for overall
  survival prediction, in: Brainlesion: Glioma, Multiple Sclerosis, Stroke and
  Traumatic Brain Injuries: 6th International Workshop, BrainLes 2020, Held in
  Conjunction with MICCAI 2020, Lima, Peru, October 4, 2020, Revised Selected
  Papers, Part II 6, Springer, 2021, pp. 215--227.

\bibitem{greene2018behavioral}
D.~J. Greene, J.~M. Koller, J.~M. Hampton, V.~Wesevich, A.~N. Van, A.~L.
  Nguyen, C.~R. Hoyt, L.~McIntyre, E.~A. Earl, R.~L. Klein, et~al., Behavioral
  interventions for reducing head motion during mri scans in children,
  Neuroimage 171 (2018) 234--245.

\bibitem{Bakas2017}
S.~Bakas, H.~Akbari, o.~Sotiras, {Advancing The Cancer Genome Atlas glioma MRI
  collections with expert segmentation labels and radiomic features},
  Scientific Data 4 (2017) 170117.

\bibitem{bakas2018identifying}
S.~Bakas, M.~Reyes, A.~Jakab, S.~Bauer, M.~Rempfler, A.~Crimi, R.~T. Shinohara,
  C.~Berger, S.~M. Ha, M.~Rozycki, et~al., Identifying the best machine
  learning algorithms for brain tumor segmentation, progression assessment, and
  overall survival prediction in the brats challenge, arXiv preprint
  arXiv:1811.02629.

\bibitem{sahayam2022brain}
S.~Sahayam, R.~Nenavath, U.~Jayaraman, S.~Prakash, Brain tumor segmentation
  using a hybrid multi resolution u-net with residual dual attention and deep
  supervision on mr images, Biomedical Signal Processing and Control 78 (2022)
  103939.

\bibitem{ronneberger2015u}
O.~Ronneberger, P.~Fischer, T.~Brox, U-net: Convolutional networks for
  biomedical image segmentation, in: Medical Image Computing and
  Computer-Assisted Intervention--MICCAI 2015: 18th International Conference,
  Munich, Germany, October 5-9, 2015, Proceedings, Part III 18, Springer, 2015,
  pp. 234--241.

\bibitem{milletari2016v}
F.~Milletari, N.~Navab, S.-A. Ahmadi, V-net: Fully convolutional neural
  networks for volumetric medical image segmentation, in: 2016 Fourth
  International Conference on 3D Vision (3DV), IEEE, 2016, pp. 565--571.

\bibitem{oktay2018attention}
O.~Oktay, J.~Schlemper, L.~L. Folgoc, M.~Lee, M.~Heinrich, K.~Misawa, K.~Mori,
  S.~McDonagh, N.~Y. Hammerla, B.~Kainz, et~al., Attention u-net: Learning
  where to look for the pancreas, arXiv preprint arXiv:1804.03999.

\bibitem{huang2020unet}
H.~Huang, L.~Lin, R.~Tong, H.~Hu, Q.~Zhang, Y.~Iwamoto, X.~Han, Y.-W. Chen,
  J.~Wu, Unet 3+: A full-scale connected unet for medical image segmentation,
  in: ICASSP 2020-2020 IEEE International Conference on Acoustics, Speech and
  Signal Processing (ICASSP), IEEE, 2020, pp. 1055--1059.

\bibitem{cao2023swin}
H.~Cao, Y.~Wang, J.~Chen, D.~Jiang, X.~Zhang, Q.~Tian, M.~Wang, Swin-unet:
  Unet-like pure transformer for medical image segmentation, in: Computer
  Vision--ECCV 2022 Workshops: Tel Aviv, Israel, October 23--27, 2022,
  Proceedings, Part III, Springer, 2023, pp. 205--218.

\bibitem{zhang2020exploring}
D.~Zhang, G.~Huang, Q.~Zhang, J.~Han, J.~Han, Y.~Wang, Y.~Yu, Exploring task
  structure for brain tumor segmentation from multi-modality mr images, IEEE
  Transactions on Image Processing 29 (2020) 9032--9043.

\bibitem{jiang2021novel}
M.~Jiang, F.~Zhai, J.~Kong, A novel deep learning model ddu-net using edge
  features to enhance brain tumor segmentation on mr images, Artificial
  Intelligence in Medicine 121 (2021) 102180.

\bibitem{zhu2023brain}
Z.~Zhu, X.~He, G.~Qi, Y.~Li, B.~Cong, Y.~Liu, Brain tumor segmentation based on
  the fusion of deep semantics and edge information in multimodal mri,
  Information Fusion 91 (2023) 376--387.

\bibitem{njeh20153d}
I.~Njeh, L.~Sallemi, I.~B. Ayed, K.~Chtourou, S.~Lehericy, D.~Galanaud, A.~B.
  Hamida, 3d multimodal mri brain glioma tumor and edema segmentation: a graph
  cut distribution matching approach, Computerized Medical Imaging and Graphics
  40 (2015) 108--119.

\bibitem{abdel2015brain}
E.~Abdel-Maksoud, M.~Elmogy, R.~Al-Awadi, Brain tumor segmentation based on a
  hybrid clustering technique, Egyptian Informatics Journal 16~(1) (2015)
  71--81.

\bibitem{dvorak2015automated}
P.~Dvorak, K.~Bartusek, W.~Kropatsch, Z.~Sm{\'e}kal, Automated multi-contrast
  brain pathological area extraction from 2d mr images, Journal of applied
  research and technology 13~(1) (2015) 58--69.

\bibitem{pinto2018hierarchical}
A.~Pinto, S.~Pereira, D.~Rasteiro, C.~A. Silva, Hierarchical brain tumour
  segmentation using extremely randomized trees, Pattern Recognition 82 (2018)
  105--117.

\bibitem{havaei2017brain}
M.~Havaei, A.~Davy, D.~Warde-Farley, A.~Biard, A.~Courville, Y.~Bengio, C.~Pal,
  P.-M. Jodoin, H.~Larochelle, Brain tumor segmentation with deep neural
  networks, Medical image analysis 35 (2017) 18--31.

\bibitem{wang20213d}
L.~Wang, J.~Du, A.~Gholipour, H.~Zhu, Z.~He, Y.~Jia, 3d dense convolutional
  neural network for fast and accurate single mr image super-resolution,
  Computerized Medical Imaging and Graphics 93 (2021) 101973.

\bibitem{kamnitsas2017efficient}
K.~Kamnitsas, C.~Ledig, V.~F. Newcombe, J.~P. Simpson, A.~D. Kane, D.~K. Menon,
  D.~Rueckert, B.~Glocker, Efficient multi-scale 3d cnn with fully connected
  crf for accurate brain lesion segmentation, Medical image analysis 36 (2017)
  61--78.

\bibitem{feng2020brain}
X.~Feng, N.~J. Tustison, S.~H. Patel, C.~H. Meyer, Brain tumor segmentation
  using an ensemble of 3d u-nets and overall survival prediction using radiomic
  features, Frontiers in computational neuroscience 14 (2020) 25.

\bibitem{kamnitsas2018ensembles}
K.~Kamnitsas, W.~Bai, E.~Ferrante, S.~McDonagh, M.~Sinclair, N.~Pawlowski,
  M.~Rajchl, M.~Lee, B.~Kainz, D.~Rueckert, et~al., Ensembles of multiple
  models and architectures for robust brain tumour segmentation, in:
  Brainlesion: Glioma, Multiple Sclerosis, Stroke and Traumatic Brain Injuries:
  Third International Workshop, BrainLes 2017, Held in Conjunction with MICCAI
  2017, Quebec City, QC, Canada, September 14, 2017, Revised Selected Papers 3,
  Springer, 2018, pp. 450--462.

\bibitem{kamnitsas2016deepmedic}
K.~Kamnitsas, E.~Ferrante, S.~Parisot, C.~Ledig, A.~V. Nori, A.~Criminisi,
  D.~Rueckert, B.~Glocker, Deepmedic for brain tumor segmentation, in:
  Brainlesion: Glioma, Multiple Sclerosis, Stroke and Traumatic Brain Injuries:
  Second International Workshop, BrainLes 2016, with the Challenges on BRATS,
  ISLES and mTOP 2016, Held in Conjunction with MICCAI 2016, Athens, Greece,
  October 17, 2016, Revised Selected Papers 2, Springer, 2016, pp. 138--149.

\bibitem{long2015fully}
J.~Long, E.~Shelhamer, T.~Darrell, Fully convolutional networks for semantic
  segmentation, in: Proceedings of the IEEE conference on computer vision and
  pattern recognition, 2015, pp. 3431--3440.

\bibitem{myronenko20193d}
A.~Myronenko, 3d mri brain tumor segmentation using autoencoder regularization,
  in: Brainlesion: Glioma, Multiple Sclerosis, Stroke and Traumatic Brain
  Injuries: 4th International Workshop, BrainLes 2018, Held in Conjunction with
  MICCAI 2018, Granada, Spain, September 16, 2018, Revised Selected Papers,
  Part II 4, Springer, 2019, pp. 311--320.

\bibitem{isensee2019no}
F.~Isensee, P.~Kickingereder, W.~Wick, M.~Bendszus, K.~H. Maier-Hein, No
  new-net, in: Brainlesion: Glioma, Multiple Sclerosis, Stroke and Traumatic
  Brain Injuries: 4th International Workshop, BrainLes 2018, Held in
  Conjunction with MICCAI 2018, Granada, Spain, September 16, 2018, Revised
  Selected Papers, Part II 4, Springer, 2019, pp. 234--244.

\bibitem{liu2022shape}
X.~Liu, Y.~Hu, J.~Chen, K.~Li, Shape and boundary-aware multi-branch model for
  semi-supervised medical image segmentation, Computers in Biology and Medicine
  143 (2022) 105252.

\bibitem{ganzetti2016quantitative}
M.~Ganzetti, N.~Wenderoth, D.~Mantini, Quantitative evaluation of intensity
  inhomogeneity correction methods for structural mr brain images,
  Neuroinformatics 14~(1) (2016) 5--21.

\bibitem{ioffe2015batch}
S.~Ioffe, C.~Szegedy, Batch normalization: Accelerating deep network training
  by reducing internal covariate shift, in: International conference on machine
  learning, PMLR, 2015, pp. 448--456.

\bibitem{gonzalez2009digital}
R.~C. Gonzalez, Digital image processing, Pearson Education, India, 2009.

\bibitem{naser2019three}
I.~Naser, J.~Al-Anssari, A.~Ralescu, Three-dimensional gradient-based laplacian
  spatial filter of a field of vectors for geometrical edges magnitude
  detection in point cloud surfaces, in: 2019 Joint 8th International
  Conference on Informatics, Electronics \& Vision (ICIEV) and 2019 3rd
  International Conference on Imaging, Vision \& Pattern Recognition (icIVPR),
  IEEE, 2019, pp. 354--361.

\bibitem{rodriguez2018beyond}
P.~Rodr{\'\i}guez, M.~A. Bautista, J.~Gonzalez, S.~Escalera, Beyond one-hot
  encoding: Lower dimensional target embedding, Image and Vision Computing 75
  (2018) 21--31.

\bibitem{bakas2017segmentation}
S.~Bakas, H.~Akbari, o.~Sotiras, Segmentation labels and radiomic features for
  the pre-operative scans of the tcga-lgg collection, The Cancer Imaging
  Archive 286.

\bibitem{bakas2017segmentationgbm}
S.~Bakas, H.~Akbari, A.~Sotiras, M.~Bilello, M.~Rozycki, J.~Kirby, J.~Freymann,
  K.~Farahani, C.~Davatzikos, {Segmentation labels and radiomic features for
  the pre-operative scans of the TCGA-GBM collection}, The Cancer Imaging
  Archive.

\bibitem{kingma2014adam}
D.~P. Kingma, J.~Ba, Adam: A method for stochastic optimization, arXiv preprint
  arXiv:1412.6980.

\bibitem{chollet2015keras}
F.~Chollet, et~al., Keras, \url{https://keras.io} (2015).

\bibitem{keras-unet-collection}
Y.~Sha, Keras-unet-collection,
  \url{https://github.com/yingkaisha/keras-unet-collection} (2021).
\newblock \href {http://dx.doi.org/10.5281/zenodo.5449801}
  {\path{doi:10.5281/zenodo.5449801}}.

\bibitem{huttenlocher1993comparing}
D.~P. Huttenlocher, G.~A. Klanderman, W.~J. Rucklidge, Comparing images using
  the hausdorff distance, IEEE Transactions on pattern analysis and machine
  intelligence 15~(9) (1993) 850--863.

\end{thebibliography}

\end{document}